\documentclass{biophys-new}
\usepackage[utf8]{inputenc}
\usepackage{graphicx}
\usepackage[colorlinks,allcolors=cyan!70!black]{hyperref}
\usepackage{textgreek}
\usepackage[normalem]{ulem}
\usepackage{xcolor}
\usepackage{hyperref}
\usepackage{amsmath}
\usepackage{amsfonts}

\usepackage{amssymb}
\usepackage{lipsum}
\usepackage{placeins}

\title{Objective clustering protocol for single-molecule data: A lifetime vs. intensity study}%% For page header:
\runningtitle{Objective clustering protocol for single-molecule data: A lifetime vs. intensity study} 

\author[1]{Michael A.C. Lovemore}
\author[1]{Bertus van Heerden}
\author[1]{Joshua L. Botha}
\author[1,*]{Tjaart P.J. Kr\"uger}
\runningauthor{Lovemore et al.} %% For page header

\affil[1]{Department of Physics, University of Pretoria, Lynnwood Road, 
Pretoria, 0002, Gauteng, South Africa}

\corrauthor[*]{tjaart.kruger@up.ac.za}
\papertype{Article}

\begin{document}
\setcounter{secnumdepth}{0}

\begin{frontmatter}
\begin{abstract}
Single-molecule spectroscopy (SMS) is an exceptionally sensitive technique, but its inherently limited photon budget produces noisy data that can readily lead to subjective analyses, fitting errors, and reduced statistical power, obscuring true subpopulations and their dynamics. Here, we present an unbiased, objective method to cluster two-dimensional single-molecule data and demonstrate it on fluorescence lifetime--intensity correlations. The clustering method is based on Gaussian mixture modeling, with the optimal number of clusters determined through {information criteria (the Akaike and Bayesian information criteria and integrated completed likelihood) and supplemented by cluster quality metrics such as average cluster tightness and the fraction of points outside confidence ellipses, which guide the selection of statistically robust and physically meaningful clusters. The protocol was benchmarked on simulated datasets spanning clean, smeared, and noisy overlap-limited regimes, and applied to experimental data from Alexa Fluor 647 and QD 605. This approach reliably recovers relevant subpopulations even in the presence of noise and overlapping distributions, providing an objective framework for analyzing single-molecule heterogeneity, with limitations arising primarily under severe geometric overlap or extreme state-occupancy imbalance where distinct populations are no longer separable.}
\end{abstract}

\begin{sigstatement}
Single-molecule spectroscopy is uniquely capable of probing the structural dynamics, reaction kinetics, and environmental responses of individual molecules and nanoscaled objects, providing simultaneous insight into parameters such as fluorescence intensity, lifetime, and spectral behavior. However, the inherent sensitivity makes this technique highly susceptible to noise, which may obscure physically relevant subpopulations and their dynamics. Clustering is often employed to assist in identifying these subpopulations, but this is typically done on the basis of user input, which is subject to bias and limits reproducibility. Here, we present an unbiased clustering protocol that can operate on any 2D dataset, enabling consistent and objective identification of molecular subpopulations.
\end{sigstatement}
\end{frontmatter}

\section{Introduction}

The pioneering work of Moerner and Kador on single molecules (SMs) \cite{moerner1989optical} paved the way for research into several highly sensitive optical imaging and spectroscopy techniques capable of detecting nanoscale molecular dynamics. Among these are SM Förster resonance energy transfer (smFRET)~\cite{roy2008practical, uphoff2010monitoring}, time-resolved fluorescence methods that are usually based on time-correlated single photon counting (TCSPC)~\cite{gruber2018isolated, bockenhauer2011conformational}, single-particle tracking techniques~\cite{van2022real, manzo2015review}, superresolution localization imaging methods such as Photoactivated Localization Microscopy (PALM) ~\cite{hess2006ultra, betzig2006imaging} and Points Accumulation for Imaging in Nanoscale Topography (PAINT) \cite{sharonov2006wide}, SM electrophysiology techniques such as single-channel patch-clamp detection~\cite{neher1976single, harms2003probing} and nanopore electrophysiology \cite{bayley2001stochastic}, and SM force spectroscopy~\cite{yen2018improving}. SM techniques are uniquely capable of detecting physically relevant subpopulations, which represent distinct functional states that the system can transition between, information that is masked or averaged out in ensemble techniques.

Raw SM experimental data can often be multidimensional depending on the type and purpose of the measurements. Each data point in a multidimensional parameter space could display information such as conformational dynamics, reaction kinetics, diffusion properties, electronic properties, and environmental responses, and these properties are often correlated with one another. The dimensionality of the data usually scales with computational complexity and visualization limitations. As such, subsets comprising only two or three parameters at a time are typically selected. Clustering can be effectively done on these lower-dimensional distributions and serves a dual purpose in denoising the data and isolating different subpopulations, some of which may be statistically rare but interesting. These data clusters may be physically linked to different functional states, for which the switching dynamics between these clusters can easily be determined. Examples include neural spike detection in patch-clamp detection~\cite{ghaderi2018electrophysiological}, spectral clustering~\cite{kruger2017reduced}, smFRET common trajectory identification \cite{blanco2015single}, and fluorescence lifetime--intensity correlation classification \cite{schlau2015single}.

{Clustering is a central tool in SMS \cite{lemmer2016unsupervised, lin2021spectral, verzelli2022unbiased, pineda2025enhanced} for identifying recurring photophysical populations in heterogeneous datasets, yet most commonly used approaches require \textit{a priori} assumptions that limit objectivity and reproducibility. Widely used methods, such as K-means and Gaussian clustering \cite{wang2013lifetime, schlau2015single, joshi2021single}, are nominally unsupervised but remain subjective in practice, as the user must specify key parameters---most notably the desired number of clusters---which can strongly bias the outcome and interpretation. K-means further imposes restrictive assumptions of spherical clusters with equal variance and no covariance structure, conditions that are rarely met in SMS data characterized by heterogeneous noise sources, anisotropic correlations, and limited photon statistics. In addition, heuristic model-selection approaches commonly used with K-means, such as the elbow method based on the within-cluster sum of squares (WCSS) \cite{umargono2020k}, rely on objective functions that decrease monotonically with increasing cluster number, rendering elbow identification highly subjective and unreliable in ambiguous or noisy regimes typical of SMS. Gaussian mixture models (GMMs) \cite{mattana2022automated, rubbens2021phenogmm} extend Gaussian clustering by modeling the data as a probabilistic mixture of multivariate Gaussian components rather than assigning hard labels, allowing cluster shape, orientation, and covariance to be explicitly captured. Importantly, GMMs are inherently compatible with information-criterion (IC) frameworks such as the Akaike information criterion (AIC) \cite{akaike1974new,burnham2002model}, Bayesian information criterion (BIC) \cite{bouman1997cluster,neath2012bayesian}, and integrated completed likelihood (ICL) \cite{biernacki2000assessing}, which enables near-automated, statistically grounded selection of the optimal number of components. Fully Bayesian clustering \cite{rubin2015bayesian,griffie2016bayesian} approaches, including Dirichlet Process GMMs (DPGMMs, based on Bayesian inference---see, e.g., Ref. \cite{gorur2010dirichlet}) in principle eliminate the need for manual model selection by treating the number of clusters as a random variable. In practice, however, these methods are computationally expensive, sensitive to prior specification, and prone to over-partitioning noisy, low-occupancy regions. These limitations are particularly problematic for photon-limited SMS data, where diffuse background fluctuations can be spuriously promoted to physically meaningful states. Although Bayesian covariance estimation can be more stable for small samples, it is inherently biased toward the prior and may generate numerous micro-clusters, undermining interpretability and reproducibility.}

SMS is an exquisitely sensitive way to detect temporal parameters such as fluorescence brightness, polarization, spectral properties (such as peak position or spectral linewidth), lifetime, and dwell times of a particular state, and can be performed on virtually any nanoscale emitter. Examples of such emitters include quantum dots (QDs) \cite{rombach2013blinking, palstra2021intermittency, palstra2021python}, fluorescent dyes \cite{zhang2021investigating}, and dendrimer-based systems in which fluorophores can be embedded within their highly-branched polymer scaffolds~\cite {paulo2010single, kohn2001single}. These systems generally exhibit binary switching behavior in their fluorescence intensity and lifetime, characterized by a well-defined bright, long-lifetime state, and a darker, quenched state, often called ``on" and ``off" states, respectively, a phenomenon known as photoblinking or fluorescence intermittency. However, many systems are known to deviate from such a bimodal behavior. Even single, fluorophore molecules have been shown to display multiple emissive states \cite{wang2013lifetime}. For QDs, the number of accessible states depends on their type, structure, and the excitation conditions. For example, type II-IV QDs generally display bimodal lifetime--intensity distributions~\cite{palstra2021python, efros2016origin}, while perovskite QDs exhibit more continuous distributions~\cite{palstra2021intermittency,yarita2017impact}. Clustering at the SM level can be a useful tool to decipher and interpret this complex fluorescence behavior \cite{rombach2013blinking,munoz2022memory, pelton2007evidence, chu2024single}. {At the level of clustered observables, locally dense populations are well approximated by Gaussian distributions, a reasonable assumption in SMS where multiple independent noise contributions lead to approximately Gaussian local behavior via the central limit theorem.}

The presence of multiple chromophores in multichromophoric systems contributes significantly to the heterogeneity in their data space. Examples of such systems include the aforementioned dendrimers as well as conjugated polymers, azobenzenes, diarylethenes, synthetic dye aggregates, DsRed, and photonic wires. Photosynthetic light-harvesting complexes form a unique group of multichromophoric systems because of the potential biological significance of data clusters. For example, a subtle change in protein conformation may switch many light-harvesting complexes between a light-harvesting and a photoprotective state~\cite{kruger2011conformational}, and these transitions have been directly correlated with photoblinking~\cite{kruger2012controlled}. In addition, these complexes display rich heterogeneity in their lifetime--intensity distributions~\cite{schlau2015single} and spectral properties~\cite{kruger2017reduced, goldsmith2010watching}. {For example, allophycocyanin, a core subunit of the main light-harvesting complex in cyanobacteria and red algae, has been shown to exhibit six distinct emissive states and a complex relationship between their fluorescence intensity and lifetime that precludes the use of simple, traditional kinetic models \cite{goldsmith2010watching}}. 

Typically, a probability distribution is used to visually reveal relevant states of an isolated SM system \cite{schlau2015single, santoso2010characterizing, kalinin2008characterizing}. These distributions can then be clustered based on the number of observed dense regions using clustering algorithms such as GMM. Each resulting cluster provides insight into the dynamics of the distinct fluorescence states and may enable inference of information such as the dwell times and switching rates between different states. There are many methods to determine these switching rates, such as the maximum-likelihood estimation of observables \cite{schmid2016single} or inverse mean dwell time \cite{schuler2013single}, though the empirical definition of the switching rate is most universal \cite{chung2018transition}. 

We have developed a general-purpose, open-source Python-based software application for SMS data analysis, called \textit{Full SMS}~\cite{botha2024advanced}, which performs, i.a., analysis of intensity time traces using a statistically robust change-point analysis~\cite{watkins2005detection}. The software also allows statistical grouping of the resolved intensity levels through a combination of agglomerative hierarchical clustering and expectation maximization clustering \cite{watkins2005detection}, which significantly improves the fitting of intensity levels and lifetimes. {While this grouping can identify physical states, it may not be statistically sufficient as an objective means to define states in a lifetime--intensity distribution, highlighting the need for a more versatile analysis method.}

{In this work, we introduce an objective, multistate clustering protocol for SMS data and demonstrate it on two-dimensional fluorescence lifetime--intensity distributions, with the lifetime and intensity data output by \textit{Full SMS} either before or after intensity-level grouping. While we here employ the Parquet file format for efficient storage and reproducible data handling, the clustering method is software-agnostic and can be applied to any SMS dataset conforming to the required structure. The protocol also supports export in multiple formats. Whereas the earlier work ~\cite{botha2024advanced} used statistical grouping primarily descriptively, the present study formalizes clustering as a statistically guided inference problem by systematically combining GMMs with multiple IC and auxiliary quality metrics, enabling objective identification of dominant photophysical populations across complex datasets. To assess robustness and practical limits, we perform controlled simulations spanning different state constructions, blinking models, and noise contributions, followed by application to two experimental datasets, viz., a rapidly blinking, randomly oriented Alexa Fluor 647 molecules dataset, as well as a small-sized QD 605 dataset \cite{munoz2022memory}. The clustering protocol is publicly available via GitHub at \url{https://github.com/BioPhysicsUP/Clustering-Protocol}}.

\section{Materials and Methods}

\subsection*{Experimental Setup}
The experimental setup is described in Ref.~\cite{kyeyune2019strong}, with some minor modifications. Briefly, a ps pulsed supercontinuum laser (Fianium, SC400-4-PP), with a repetition rate of $40$ MHz, was used as the illumination source, where the central wavelength of excitation was selected using an acousto-optical tunable filter (Crystal Technology). The excitation beam was circularly polarized before passing through a spatial filter to isolate the $\text{TEM}_{00}$ mode, and thereafter focused into a near-diffraction-limited spot by a 1.45-NA Nikon oil-immersion objective in an epifluorescent configuration. The fluorescence photons were captured by the same objective, focused through a confocal pinhole using an appropriately selected dichroic beamsplitter, spectrally cleaned using a suitable long-pass filter, and detected by a single-photon avalanche photodiode (PD-050-CTE, Micro Photon Devices, Bolzano, Italy, IRF $\sim128$ ps). 

\subsection*{Sample Preparation}
Alexa Fluor 647 (Alexa hereafter; Thermo Fisher Scientific) 
was diluted in a solution of $6$ mM 2-(N-morpholino)ethanesulfonic acid (MES) buffer (pH 7) to $\sim$5 pM. No redox chemicals were added. This solution, which also contained $4$\% (w/w) polyvinyl alcohol (PVA), was spin-coated onto a glass microscope coverslip before measurement. Individual molecules were identified from a raster scan and excited using 200 nW of 633 nm light. For this sample, the dichroic mirror was FF650-Di01-25$\times$36 (Semrock) and the fluorescence filter was FELH0650 (Thorlabs).

QD 605 ITK carboxyl-derivatized QDs (Thermo Fisher Scientific) were diluted 
in a solution of $10$ mM MES buffer (pH 7) with $0.2$ mM $\text{MgCl}_2$ and $0.05$\% (w/v) Tween-20 to a concentration of $\sim80$ pM and sandwiched between two microscope coverslips, with the bottom one treated with poly-L-lysine (PLL). The QDs were investigated one by one using $488$ nm excitation at $140$ nW, a 605dcxt dichroic beamsplitter (Chroma Technology), and a 600LPF fluorescence filter (Edmund Optics).

\subsection*{Simulations}
Simulated datasets were generated to benchmark the clustering algorithm, the code of which is available at \url{https://github.com/BioPhysicsUP/SMS-Simulations}. The inherent noise in SMS experiments includes the fundamental shot noise (obeying Poisson statistics), detector noise (obeying Gaussian statistics), and ambient background noise. The latter arises from stray light sources such as scattering and autofluorescence from optical components and appears as a constant background level, where each stray light source is also affected by shot noise. Background noise is especially problematic for the lowest-intensity populations, where it significantly reduces the signal-to-noise ratio. The detector noise includes dark counts and readout noise, but can be accounted for by taking a background measurement and subtracting that level from the intensity data. The shot noise is inherent due to the discrete nature of photons arriving at random times. This is generally the major noise contributor in experiments and cannot be subtracted. To simulate the data, random lifetime--intensity populations were generated by defining a sampling population with a user-specified mean and standard deviation (SD) for the lifetime and intensity of each state. {Dwell times were then generated on a per-particle basis. In the main text, both quenched (“off”) and unquenched (“on”) states were drawn from exponential distributions (see Eq. \eqref{expplaw}). In the SI, we explored more complex scenarios in which the “off” dwell times follow a power-law distribution, while the “on” dwell times are drawn from either an exponential, a simple power-law (Eq. \eqref{powerlaw}), or a truncated power-law (Eq. \eqref{trunpowerlaw}). This demonstrates the capability of the clustering protocol to handle datasets governed by different underlying statistical models.}

{The exponential dwell-time probability distribution function is given by}

{
\begin{equation}\label{expplaw}
    P(t)=\frac{1}{\tau}e^{-t/\tau},
\end{equation}}

\noindent {where $t$ is the dwell time and $\tau$ is a sampled mean dwell time, set to $\tau=2$ s for the off-state and $\tau=4$ s for the on-states in this study. } {The simple power-law probability distribution is given by}

{
\begin{equation}\label{powerlaw}
    P(t)=t^{-\alpha},
\end{equation}}

\noindent {where $\alpha$ is the power-law constant, set to $\alpha=1.2$ for this study.} The truncated power-law function is given by

\begin{equation}\label{trunpowerlaw}
    P(t)\propto t^{-\alpha} \cdot e^{\frac{-t}{\tau_c}},
\end{equation}

\noindent where $\tau_c$ acts as the exponential cutoff time after which the exponential decay starts to dominate. The truncated power law is characterized by a high probability of short dwell-time events and a low probability of long dwell times. For this particular study, we set $\alpha=1.2$ and $\tau_c=20$ s. 

{For each particle, dwell times were drawn from these distributions}, where the corresponding intensities and lifetimes were sampled from the defined means and SDs, thereby simulating states with normally distributed noise. Finally, Poisson noise (representing shot noise) was added to the intensities to yield {the desired simulated intensity traces.}

For molecules with a distinct dipole (such as Alexa), the excitation probability depends strongly on the dipole orientation with respect to the polarization of the excitation light. In the dipole approximation, this probability is defined by

\begin{equation}
\label{pexc}
    P_{\text{Exc}} \propto |\vec{E}\cdot\vec{\mu}|^2 \propto \sin^2{\theta}(\cos\phi + \epsilon\sin\phi)^2,
\end{equation}

\noindent where $\theta \in [0,\pi]$ and $\phi \in [0,2\pi]$ are, respectively, the polar and azimuthal angles of the molecule's transition dipole moment, and $\epsilon$ is the ellipticity of the excitation light's polarization ellipse. The fluorescence quantum yield was assumed to be unaffected by the light polarization and molecule orientation.

\subsection*{Data Analysis}\label{datanalysis}

The experimental data were recorded in time-tagged time-resolved (TTTR) mode, stored in HDF5 format, and analyzed using \textit{Full SMS}~\cite{botha2024advanced}. Data screening was the first step, whereby particles bleached (visible as an irreversible loss of fluorescence) during the raster scan were omitted, photon bursts were removed, and only the brightest particles were selected. The time traces of the remaining particles were then analyzed within a specific time region until partial or complete photodegradation was observed. The fluorescence intensity levels were then resolved and subsequently grouped using the software's native clustering functionality, which provided the statistically most probable levels. The fluorescence lifetime of each grouped intensity level was determined using $\chi^2$ and Durbin-Watson \cite{durbin1950testing} goodness-of-fit parameters.

The correlated lifetime--intensity data output of the \textit{Full SMS }software was further analyzed in a raw Python environment to perform additional data clustering. This clustering algorithm (hereafter referred to as the clustering protocol) objectively determines the optimal number of identifiable populations (states) to which a single particle (object) can switch. Data were clustered using GMM, assuming that all points in the distribution were drawn from a mixture of Gaussian distributions with unknown parameters. For each GMM cluster, a mean vector and a covariance matrix were estimated, representing the cluster center and the shape/spread of the data. Data points were assigned according to the highest likelihood of belonging to a cluster. The optimal number of GMM clusters was determined using {three IC metrics---AIC, BIC, and ICL---supplemented by cluster-quality measures including relative IC gain, average cluster tightness, and the fraction of points outside confidence regions. This combination enables reproducible and minimally biased selection of the number of clusters.}

{The AIC score was calculated as}

\begin{equation}
    \text{AIC}=2k-2\cdot\text{ln}(L),
\end{equation}

\noindent {where $k$ is the number of free parameters and $L$ the maximized likelihood. AIC penalizes model complexity weakly and therefore tends to favor models with a larger number of clusters, emphasizing predictive accuracy. The BIC score, in comparison, is defined as}

\begin{equation}
    \text{BIC} = k\cdot \text{ln}(n)-2\cdot\text{ln}(L),
\end{equation}

\noindent  {where $n$ is the number of data points. BIC penalizes model complexity more strongly than AIC, favoring parsimonious solutions consistent with Occam's razor. ICL extends the BIC formulation by explicitly incorporating cluster assignment certainty:}

\begin{equation}
    \text{ICL} = \text{BIC} - \sum_{i=1}^{n}\sum_{j=1}^{k} \tau_{ij}\cdot \ln(\tau_{ij}),
\end{equation}

\noindent {where $\tau_{ij}$ denotes the posterior probability that data point $i$ belongs to cluster $j$. The additional entropy term penalizes overlapping or uncertain assignments, favoring well-separated clusters. Although BIC is most commonly used in SMS studies, including AIC and ICL provides complementary information on predictive accuracy, model parsimony, and cluster separation. All three IC metrics are derived from the GMM likelihood framework and therefore integrate naturally with probabilistic cluster modeling.}

{IC scores were computed for each candidate number of GMM clusters. The conventional approach is to identify the optimal solution as the model yielding a global minimum IC value. However, when these flexible mixture models are applied to noisy SM data, which is subject to heterogeneity, the IC values frequently decrease with increasing cluster number, and reach a global minimum at unrealistically large numbers of components, leading to possible overpartitioning and substantial computational cost, while offering little physical interpretability (as illustrated in the experimental results discussed in this study). To mitigate this, our algorithm focuses on substantial improvements in model fit as quantified by decreases in the IC, rather than on absolute IC minimization. Specifically, we computed the relative change in the IC score between successive cluster numbers, normalized to the maximum observed change. Here, an IC gain is defined as a pronounced reduction in the IC value upon increasing the number of clusters, yielding an elbow point, which denotes a point at which a steeper gradient transitions to a markedly shallower gradient. Cluster numbers producing sharp IC decreases, whose gradients exceed a threshold set by one standard deviation above the mean IC gradient, were identified as prominent local elbow points; once the IC gradient falls below this threshold, that cluster is no longer considered to provide statistically meaningful improvement. These prominent local elbow points were flagged as candidate solutions, generating a set of statistically grounded cluster choices. Importantly, experimental datasets may exhibit multiple such prominent elbows, reflecting competing yet plausible descriptions of the underlying state structure. Consequently, for such datasets, IC analysis alone is insufficient to select a unique optimal solution. From the resulting candidate set, final model selection was therefore performed using complementary cluster-quality metrics (\textit{vide infra}), which quantify cluster compactness, separation, and physical plausibility. This two-stage procedure prioritizes models that achieve both statistically meaningful IC gains and physically interpretable clustering, avoids reliance on subjective visual inspection, and mitigates systematic overfitting while remaining sensitive to genuine molecular state structure.}

Ellipses were drawn around each cluster center {to represent} confidence contours for each fitted GMM component. For a two-dimensional Gaussian with mean $\vec\mu$ and positive semi-definite covariance matrix $\Sigma$, the contour is defined by the squared Mahalanobis distance:

\begin{equation}
    D^2=(\vec{x}-\vec{\mu})^T\Sigma^{-1}(\vec{x}-\vec{\mu})=\chi_2^2({p}),
\end{equation}

\noindent{where} $\vec{x}$ is a point on the lifetime--intensity plot, and $\chi^2_{2}(p)$ denotes the quantile of the chi-squared distribution with two degrees of freedom {corresponding to the confidence level} $p$. This ellipse encloses a fraction $p$ of the probability mass around the cluster center of each GMM component. 

{These confidence ellipses also serve as quantitative guides in cluster evaluation. Cluster tightness was defined as the mean squared Mahalanobis distance of points to their assigned cluster centers, averaged across all clusters ($\langle D^2 \rangle$). For an ideal two-dimensional Gaussian, the expected value is approximately equal to the dimensionality, i.e., $\langle D^2 \rangle \approx 2$. Deviations indicate overly compact ($<2$) or loose ($>2$) clusters. The fraction of points outside the confidence ellipse (FO) was also computed, with deviations from the expected value, $1-p$, indicating overlap, poor separation, or model mismatch. Together, these metrics provide intuitive checks on clustering robustness.} Throughout this study, $p$ was selected empirically to minimize cluster overlap while preserving adequate information about cluster spread. {Candidate clustering solutions were selected based on a balance between statistical parsimony, cluster tightness, and information gain across multiple metrics, minimizing subjective bias in cluster number determination.}

{Although many other clustering algorithms could be employed in principle \cite{hartigan1975clustering, xu2005survey}, careful choice of an appropriate algorithm is key. The specific data structure is an important consideration. For lifetime--intensity data, the point density varies strongly with intensity, lifetime, and photon count, causing some physically meaningful low-intensity or short-lifetime states to be misclassified as noise or fragmented across clusters in clustering paradigms such as DBSCAN. More generally, SMS observables are naturally described by elliptical distributions with non-zero covariance, rendering GMMs an appropriate clustering choice as they accommodate this structure through full covariance estimation and probabilistic assignment. We therefore used GMM as the primary clustering framework in this work, while comparing the performance of the DPGMM and K-means clustering methods on selected datasets.}

{In Bayesian formulations of GMMs such as DPGMMs, prior distributions are placed on the mixture weights and component parameters, allowing the effective number of clusters to be inferred by suppressing components with negligible posterior weight. In this study, DPGMM was implemented by specifying an initial upper bound on the number of mixture components, chosen deliberately to be slightly higher than the expected number of states, after which the model redistributes weight among clusters during inference. This reflects a realistic experimental scenario in which the true number of populations is unknown, and a robust clustering algorithm should collapse redundant components. The resulting solution depends sensitively on the choice of hyperparameters, particularly the concentration parameter, and under inappropriate prior specification, DPGMMs may assign low-occupancy or spurious components (called “ghost” clusters). These solutions were therefore evaluated with respect to both the inferred number of clusters and the emergence of such ghost components.}

{By contrast, K-means clustering was applied as a distance-based reference method. The algorithm performs hard assignments by minimizing the squared Euclidean distance (i.e., WCSS) between data points and cluster centroids, assigning each point $\vec{x}_i$ to the cluster $k$, with cluster center $\vec{\mu}_k$, such that $|\vec{x}_i - \vec{\mu}_k|^2$ is minimized. WCSS values were computed for cluster numbers ranging from one up to a predefined maximum and plotted as an elbow-point curve. The optimal number of clusters was identified as the point at which the WCSS trend transitions from a steep to a shallow gradient, determined by fitting linear trends to the low- and high-cluster-number regimes and locating their intersection. Cluster boundaries are therefore defined by loci of equal distance between centroids, resulting in linear decision surfaces corresponding to Voronoi partitions of the feature space. This formulation assumes spherical clusters with equal variance and neglects covariance between observables, properties that are generally incompatible with lifetime--intensity data exhibiting correlated noise and unequal spread. Under these conditions, distance-based hard clustering is expected to yield unstable or biased partitions relative to covariance-aware probabilistic models.}

Once the optimal number of populations was determined, {clustering performance was evaluated using the Adjusted Rand Index (ARI) \cite{hubert1985comparing}, which quantifies agreement between the true simulated state labels and the cluster assignments returned by the GMM at the level of individual dwell segments. Unlike metrics based solely on cluster-center matching, ARI directly measures point-level assignment accuracy and explicitly penalizes both merged and missing clusters, making it well-suited for situations in which the recovered number of clusters differs from the true number of physical states. An ARI value of unity indicates perfect recovery, while values approaching zero correspond to random assignment.} Other relevant information, such as the switching rates and switching frequencies between states, {could then be extracted. }

The switching rate from state $i$ to state $j$ was defined as 

\begin{equation}\label{switchingrates}
	k_{ij} = \frac{N_{ij}}{\tau_i}  , \\
\end{equation}

\noindent where $N_{ij}$ is the total number of switches from state $i$ to state $j$ and $\tau_i$ is the total dwell time in state $i$. The switching frequency indicates how frequently the particle or molecule switches between states $i$ and $j$ and was defined as 

\begin{equation}\label{switchingfreq}
	f=\frac{N_{ij}+N_{ji}}{\tau_i+\tau_j}. 
\end{equation}

The switching ratio, defined as

\begin{equation}\label{switchingratio}
	R_{ij} = \frac{k_{ij}}{k_{ji}},
\end{equation}

\noindent gives the ratio of the forward to reverse switching rates and is a measure of the state preference. We used the condition $i<j$, in which case $R_{ij}<1$ indicates the sample prefers switching to the lower intensity state, $i$, and $R_{ij}>1$ indicates switching to the higher intensity state, $j$, is favored.

To make the clustering method more robust, we incorporated a modified interquartile range (IQR) outlier detection rule, which excludes data that lie sufficiently far away from the corresponding cluster center. Specifically, if the distance between the data point coordinates and the cluster center coordinates satisfies the inequality

\begin{equation}\label{IQRrule}
    \text{distance}>\text{median} + k\cdot \text{IQR},
\end{equation} 

\noindent for some constant $k$, then the data was excluded. Following the standard convention introduced by Tukey's boxplot rule, we set $k=1.5$.

\section{Results}
\subsection*{Simulated Data}

{Simulated datasets were generated from fluorescence lifetime--intensity models with varying numbers of underlying states to evaluate the clustering protocol under controlled conditions, with all analyses performed at a confidence level of $0.95$. Background noise was included in all datasets, and on- and off-state dwell times were drawn from exponential distributions. Alternative dwell-time models are presented in the SI. Normally distributed noise around each state’s mean, constrained by its standard deviation, was added to mimic dense SMS populations. To systematically assess the protocol’s limits, we started with simple, well-defined systems comprising equally populated, tight clusters (i.e., small sampling SDs) and progressively introduced additional complexity. This approach allows identification of the approximate point at which increasing overlap and diffuse populations compromise cluster recovery, thereby defining the “breaking point” of the algorithm. At selected points in this analysis, clustering performance was additionally compared against DPGMM, in which the effective number of components was inferred within a Bayesian framework, and K-means clustering, with the number of clusters determined using the elbow criterion.}\\

\noindent {\textbf{Well-defined system with equal probability states}}\\

\begin{figure}[h!tbp]
    \centering
    \includegraphics[width=0.7\linewidth]{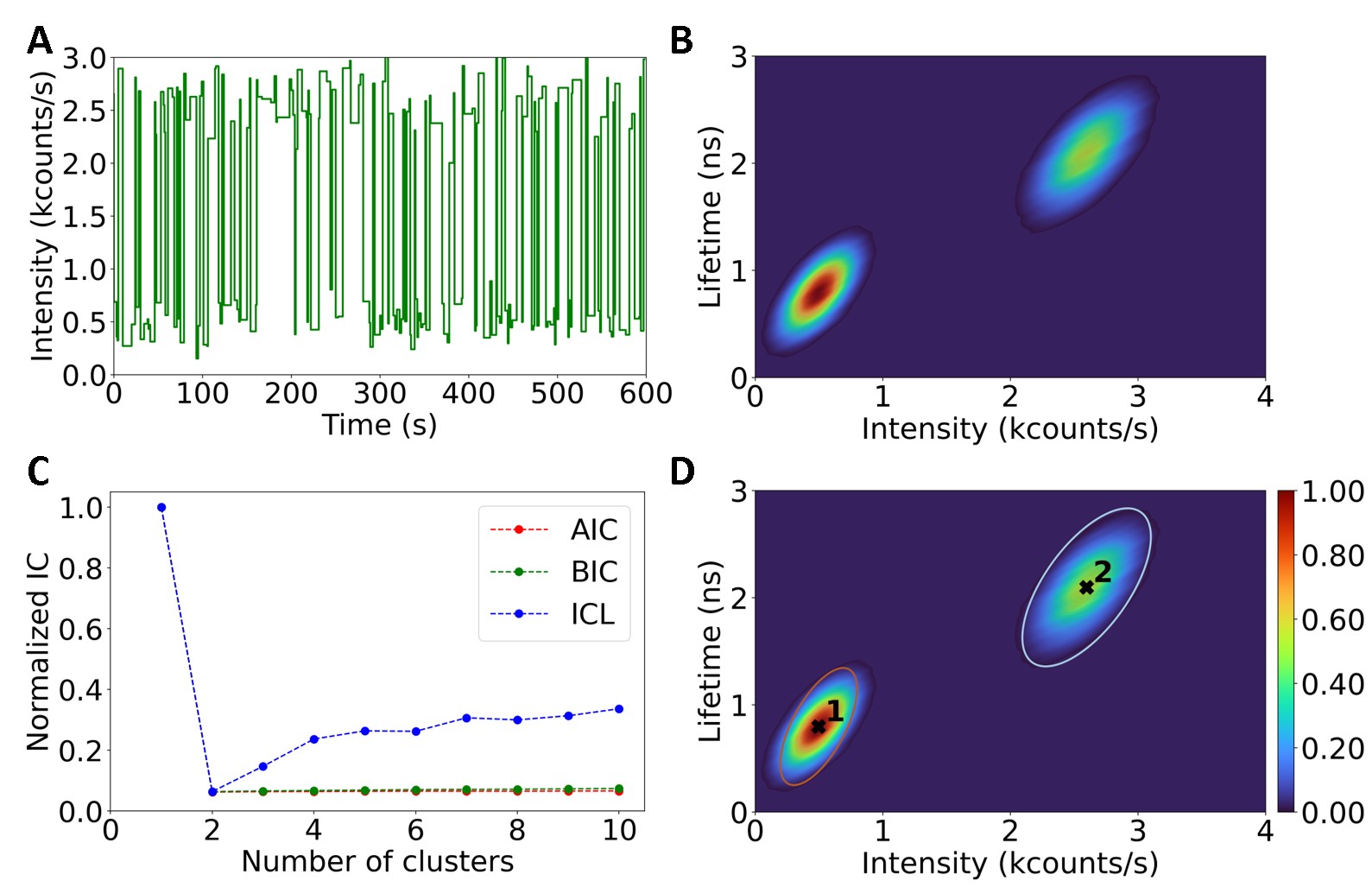}
    \caption{{Results of the clustering protocol applied to a two-state simulated fluorescence lifetime--intensity dataset (10-min time, 300 particles). A. Representative intensity trace (40-ms binning shown) illustrating rapid photoblinking between emissive and background states. B. Lifetime--intensity distribution, showing two distinct populations. C. IC score plots, indicating two clusters as optimal, with $\langle D^2\rangle = 2$ and FO $= 0.05$. D. Optimally clustered distribution with cluster centers (black crosses) and 0.95 confidence ellipses; cluster centers at (500 cps, 0.8 ns) and (2601 cps, 2.1 ns), ARI $=1$.}}

    \label{twostatesim}
\end{figure}

\noindent {To simulate a two-state system (Fig. \ref{twostatesim}), intensities were sampled at $500 \pm 120$ cps and $2600 \pm 200$ cps, where the values denote the mean $\pm$ SD from normal distributions, while lifetimes were correspondingly sampled at $0.8 \pm 0.22$ ns and $2.1 \pm 0.3$ ns, respectively. For a four-state system (Fig. \ref{fourstatesim}), two additional states were introduced with intensities of $1200 \pm 240$ cps and $3100 \pm 200$ cps, and corresponding lifetimes of $2.4 \pm 0.2$ ns and $1.1 \pm 0.2$ ns, respectively. Under these conditions, the clustering protocol is expected to recover well-defined populations that match the underlying model. Figure \ref{twostatesim}A shows a representative simulated intensity trace exhibiting rapid photoblinking, where the off-state corresponds to the background noise intensity. The lifetime--intensity distribution (Fig. \ref{twostatesim}B) clearly shows two populations. The IC score plots (AIC, BIC, and ICL; Fig. \ref{twostatesim}C) exhibit a sharp decrease to two clusters, with negligible improvement beyond these numbers. ICL deviates significantly from AIC and BIC beyond two clusters due to its additional penalty for cluster overlap. Cluster tightness, defined by the mean Mahalanobis distance, matched the expected value of $\langle D^2\rangle=2$, and the average fraction of points outside the $0.95$ confidence ellipse was FO $= 0.05$, confirming that the clusters are compact and accurately reflect the underlying populations. Figure \ref{twostatesim}D shows the optimally clustered two-state system, with cluster centers at $(500 \text{ cps}, 0.8 \text{ ns})$ and $(2601 \text{ cps}, 2.1 \text{ ns})$, and ARI $=1$, indicating perfect ground-truth recovery. The clustering of the additional well-defined four-state system, which yielded identical results ($\langle D^2\rangle=2$, FO = 0.05, ARI $=1$), shown in Figure \ref{fourstatesim}, indicates that higher-order well-defined states do not affect recovery of the simpler two-state clusters.}\\

\noindent {As a first comparison to alternative clustering approaches, Figure~\ref{compareclustermethodone} shows the application of DPGMM and K-means clustering to the same well-defined two-state dataset. For the DPGMM analysis, the number of prior components was set to three with a Dirichlet concentration parameter of 0.1, such that one component would be expected to acquire negligible weight and be suppressed. This deliberate over-specification reflects a realistic experimental scenario in which the true number of states is unknown, and a robust Bayesian method should collapse redundant components rather than retain spurious ones. Instead, the DPGMM method identified three clusters as optimal. When projected onto the lifetime--intensity distribution (Fig.~\ref{compareclustermethodone}A), two inferred clusters were assigned to the same high-intensity state with nearly coincident centers, indicating the formation of a low-occupancy spurious (“ghost”) cluster. Although the cluster centers themselves align well with the underlying populations, this redundant partitioning highlights the sensitivity of DPGMM solutions to prior specification and concentration parameters, with identification of the artifact requiring explicit visualization of the clustering output. The computational cost of the DPGMM analysis was also approximately twice that of the standard GMM protocol for this simple case. For comparison, K-means clustering was applied via the elbow-point selection (Fig.~\ref{compareclustermethodone}B), which clearly identified two clusters as optimal. The resulting clustering solution (Fig.~\ref{compareclustermethodone}C) correctly separated the two populations, yielding visually reasonable partitions, albeit with a computational time again approximately twice that of our GMM-based protocol. While K-means performs adequately in this trivial, well-separated scenario, its distance-based hard assignments and restrictive cluster-shape assumptions limit its general applicability, whereas the DPGMM comparison already reveals sensitivity to priors even under idealized conditions.}\\

\noindent {\textbf{Well-defined two-state system using different blinking models}}\\

\noindent {Additional blinking models are presented in Figure \ref{SIotherdwells}, where off-state dwell times follow a power-law distribution and on-state dwell times follow power-law (Fig. \ref{SIotherdwells}A), exponential (Fig. \ref{SIotherdwells}C), or truncated power-law (Figs. \ref{SIotherdwells}E, G) statistics, where for the latter, a moderate power-law constant of 1.2 (Fig. \ref{SIotherdwells}E) and a very high value of 3.5 (Fig. \ref{SIotherdwells}G) were investigated. The corresponding clustering results (Figs.~\ref{SIotherdwells}B, D, F, and H, respectively) are practically identical, each showing two clusters as the optimal choice ($\langle D^2\rangle=2$, FO = 0.05) and yielding perfect ground-truth recovery (ARI = 1). This suggests that variations in blinking statistics had no effect on the accuracy of state recovery for this simple two-state system.}\\

\noindent {\textbf{Polarization and dipole orientation effects}}\\

\begin{figure}[h!tbp]
    \centering
    \includegraphics[width=0.7\linewidth]{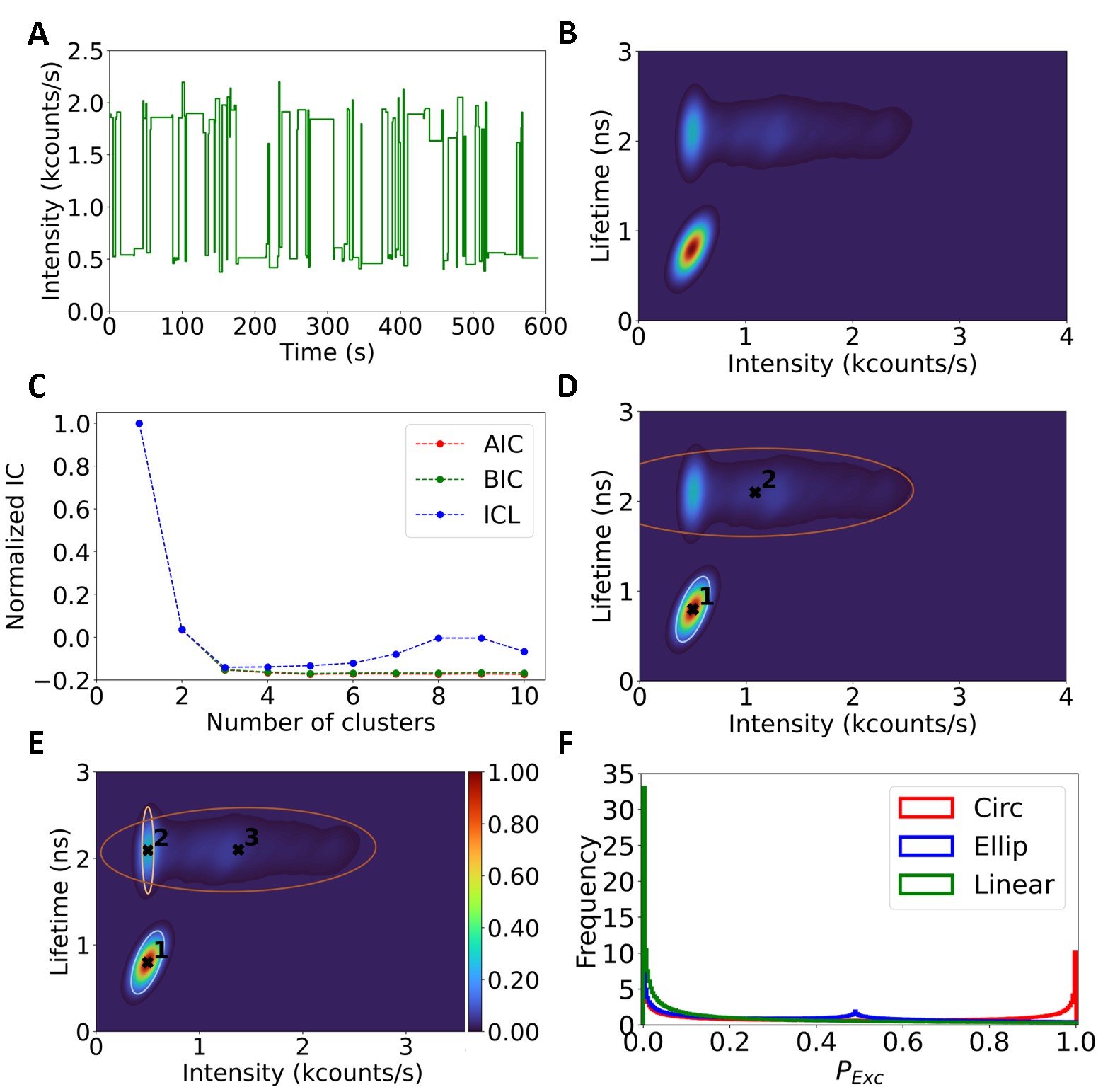}
    \caption{Influence of the excitation light polarization on the simulated intensity of a molecular dipole. A. Example intensity trace (with 40 ms binning) for circularly polarized light and a dipole orientation of $\theta=\frac{2\pi}{3}$ and $\phi=\frac{3\pi}{4}$, simulated for a time of 10 min. B. Lifetime--intensity distribution of a set of 300 fixed but randomly oriented dipoles. {C. Corresponding IC score plots, with two ($\langle D^2\rangle = 2$, FO $= 0.047$) and three clusters ($\langle D^2\rangle = 1.987$, FO $= 0.045$) identified as optimal choices. D. Clustering result for the two-cluster solution, with centers at (500 cps, 0.8 ns) and (1102 cps, 2.1 ns) shown as black crosses with 0.95 confidence ellipses; ARI = 1. E. Three-cluster solution, centers at (500 cps, 0.8 ns), (500 cps, 2.1 ns), and (1650 cps, 2.1 ns); ARI = 0.8, indicating some misassignment due to the broadened distribution of the higher-intensity population. F. Histogram of excitation probabilities for $10^6$ randomly oriented dipoles under circular ($\epsilon=1$, red), elliptical ($\epsilon=0.7$, blue), and linear ($\epsilon=0$, green) polarization.}}
    \label{smearedS1}
\end{figure}

\noindent To incorporate the effect of excitation light polarization and particle orientation, we applied Eq. \eqref{pexc} to the {higher-intensity state from Figure \ref{twostatesim}B} for circularly polarized light ($\epsilon=1$) with randomized dipole orientations. {For circularly polarized excitation, the excitation probability depends only on the polar angle $\theta$, following a $\sin^2\theta$ dependence.} {This introduces broadening of the population toward lower fluorescence intensities.} An example intensity trace, for a particle assumed to have a fixed orientation defined by $\theta=\frac{2\pi}{3}$, $\phi=\frac{3\pi}{4}$, yielding an excitation probability of $P_\text{{Exc}}=0.75$ according to Eq. \eqref{pexc}, is shown in Figure \ref{smearedS1}, where photoblinking occurs between the off-state and the reduced on-state corresponding to a simulated maximum emission intensity of {1950 cps}. The smeared lifetime--intensity distribution for 300 randomly oriented particles {(Fig. \ref{smearedS1}B) shows strong broadening of the unquenched state down to the background level, with an average background-subtracted intensity of $867$ cps, consistent with the expected $1/3$ of the maximum intensity.}

Applying the clustering protocol to this smeared dataset {yielded the IC score plots in Figure \ref{smearedS1}C, where the elbow points, corresponding to the points of greatest gradient increase and relative IC gain, are at two and three clusters for all three IC metrics. The result for two clusters (with $\langle D^2\rangle = 2$ and FO = 0.047) is shown in Figure \ref{smearedS1}D, where the cluster of the higher state is centered offset to its mean intensity (1102 cps, 2.1 ns), and that of the lower state exactly coincides with the densely populated off-state (500 cps, 0.8 ns).} {Despite the density-weighted displacement of the second cluster’s intensity, ARI $=1$, indicating perfect point-level recovery per cluster.}

{The three-cluster solution (Fig. \ref{smearedS1}E, with $\langle D^2\rangle = 1.987$ and FO = 0.045) splits the long-lifetime data into two clusters, with cluster 2 centered at the dense background population (500 cps, 2.1 ns) and cluster 3 at the mean of the remaining data (1650 cps, 2.1 ns).} Although cluster 3 should align with 2600 cps, the lower density of points (caused by broadening) at this intensity shifts the cluster center toward the denser region at intermediate intensities. For this solution, {ARI $=0.8$, showing that most points are correctly assigned, and the differences in $\langle D^2\rangle$ (1.987) and FO (0.045) relative to the two-cluster solution reflect only the broadening of the higher-intensity population. These metrics support the conclusion that two clusters (which correspond to ground truth) represent the optimal solution when considering both statistical criteria and cluster interpretability, even though the IC score for three clusters is slightly lower.}

Notably, any other ellipticity of the excitation light’s polarization ellipse gives rise to very similar smearing of the intensity distributions down to the background level. In this representation, the ellipticity merely affects the excitation probability density, shown as a histogram in Figure \ref{smearedS1}F for equally-weighted randomly oriented dipoles interacting with three different polarization states of an incident radiation field. All three cases (linear, elliptical, and circular polarization) display a peak at an excitation probability $P_{\text{Exc}} = 0$, and any non-zero choice of $\epsilon$ introduces an additional peak at $P_{\text{Exc}} = \epsilon^2$.\\

\noindent {\textbf{Five-state system: effect of state overlap}}\\

\noindent {To assess the impact of state overlap on cluster recoverability, the well-defined two-state system in Figure \ref{twostatesim} was extended by three additional states at fixed intensity--lifetime coordinates: (1200 cps, 2.4 ns), (1500 cps, 1.4 ns), and (2200 cps, 0.8 ns). The original two states were kept unchanged in both mean values and SDs. All five states were assigned equal occupation probabilities to maintain constant overall data density. To isolate the effect of overlap, only the SDs of the three added states---scaled proportionally to their mean intensity and lifetime---were systematically increased, while their mean positions and probabilities were held fixed, thereby moving from well-defined to diffuse states. This controlled broadening progressively increased the spatial extent and overlap of the three additional, diffuse populations with one another and with the original two well-defined populations, allowing direct identification of the regime in which overlap degrades cluster separability and recovery. The resulting degree of overlap was quantified using the average pairwise Gaussian overlap (APGO) metric, calculated from standard Gaussian overlap integrals between all state pairs.} 

\begin{figure}[h!tbp]
    \centering
    \includegraphics[width=1\linewidth]{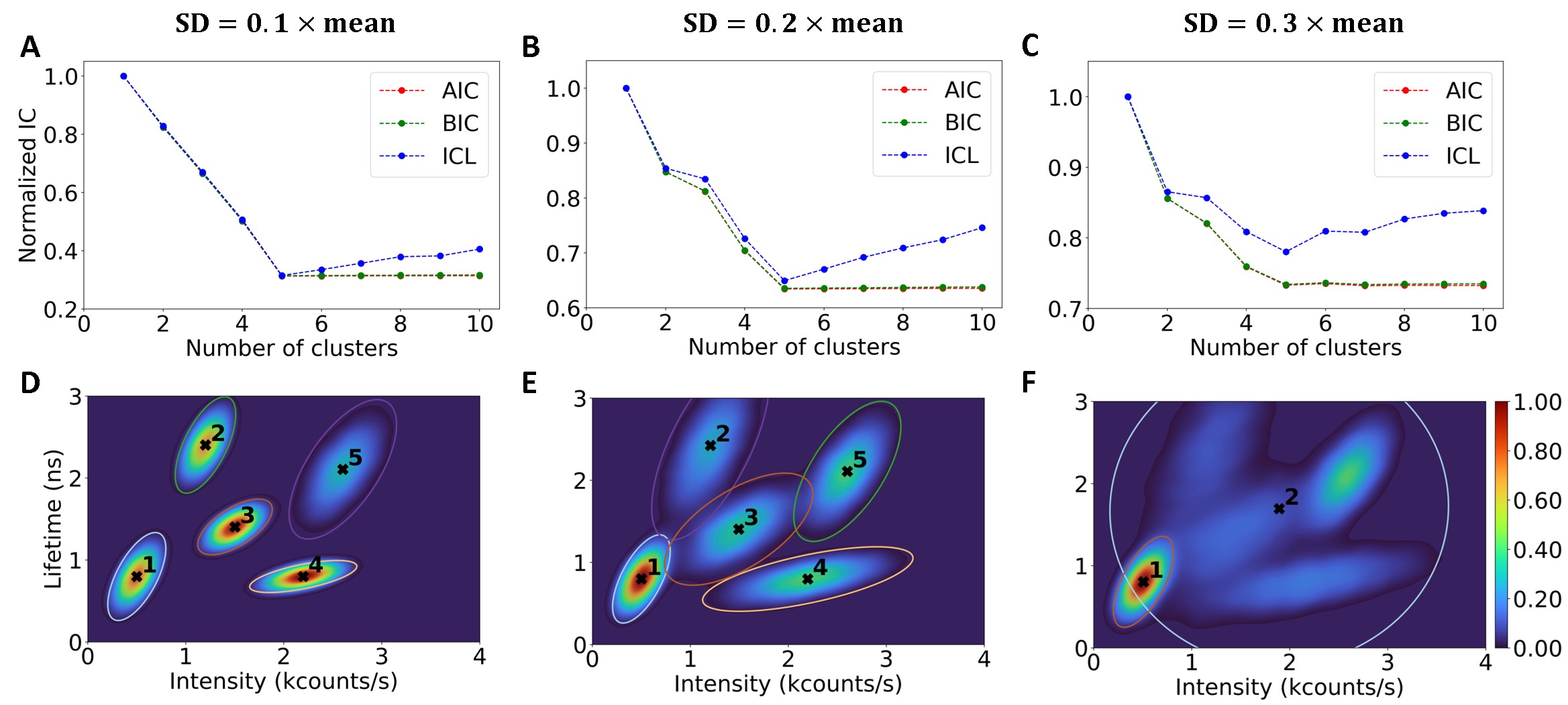}
    \caption{{Effect of state overlap by increasing state widths on five-state clustering recovery. A--C. IC score plots for each SD case: SD = 0.1 $\times$ mean (A) shows minimal overlap (APGO = 0.05), with AIC and BIC identifying five clusters ($\langle D^2\rangle = 2$, FO = 0.05); SD = 0.2 $\times$ mean (B) exhibits competing minima at two ($\langle D^2\rangle = 1.98$, FO = 0.028) and five clusters ($\langle D^2\rangle = 1.94$, FO = 0.04); SD = 0.3 $\times$ mean (C) shows substantial overlap (APGO = 0.43), with two clusters selected as optimal ($\langle D^2\rangle = 1.96$, FO = 0.03), and a formally accessible five-cluster solution showing poorer compactness ($\langle D^2\rangle = 1.8$, FO = 0.03). D--F. Corresponding clustering results: SD = 0.1 $\times$ mean (D) recovers all five clusters perfectly (ARI = 1); SD = 0.2 $\times$ mean (E) shows near-complete recovery for five clusters (ARI = 0.95) (F) recovers only two clusters (ARI = 0.2) with centers at (500 cps, 0.8 ns) and (1920 cps, 1.65 ns).}}

    \label{fivestateSDsstudy}
\end{figure}

{Figure \ref{fivestateSDsstudy} shows the effect of increasing diffuse-state widths on clustering recovery, examining the three values SD = 0.1 $\times$ mean, SD = 0.2 $\times$ mean, and SD = 0.3 $\times$ mean. For SD = 0.1 $\times$ mean (left column), APGO = 0.05, indicating minimal overlap between populations. AIC and BIC follow identical trajectories while ICL deviates only beyond five clusters (Fig. \ref{fivestateSDsstudy}A), consistently identifying five clusters as optimal with $\langle D^2\rangle = 2$ and FO = 0.05. For SD = 0.2 $\times$ mean (middle column), APGO rises to 0.24, reflecting moderate overlap. The IC curves (Fig. \ref{fivestateSDsstudy}B) exhibit competing candidate elbow points at two and five clusters. The two-cluster solution yields marginally tighter clusters, with $\langle D^2\rangle = 1.98$ and FO = 0.028, whereas the five-cluster solution is slightly less tight ($\langle D^2\rangle = 1.94$) with a higher but still moderate FO = 0.04. Importantly, the gain in tightness obtained by merging the data into two clusters is small, while the increase in FO for the five-cluster solution reflects the deliberate separation of the three additional states rather than a loss of data coverage. In this regime, the five-cluster partition captures a larger fraction of the data structure by resolving the additional states individually, with only a modest relaxation in tightness. Consequently, despite the slightly tighter two-cluster solution, the five-cluster solution provides a more faithful representation of the full dataset, balancing cluster tightness with improved state resolution. For SD = 0.3 $\times$ mean (right column), the APGO further increases to 0.43, indicating substantial population merging. All three IC metrics again identify two and five clusters as the optimal elbow points (Fig. \ref{fivestateSDsstudy}C), with corresponding auxiliary metrics of $\langle D^2\rangle = 1.96$, FO = 0.03, and $\langle D^2\rangle = 1.8$, FO = 0.3, respectively. In this case, the five-cluster solution shows markedly poorer compactness than before, while the two-cluster solution arises as the more statistically grounded solution, based on the cluster-quality metrics.}

{Figures \ref{fivestateSDsstudy} D--F show the corresponding clustered results. For SD = 0.1 $\times$ mean (Fig. \ref{fivestateSDsstudy}D), five clusters are perfectly recovered with cluster centers exactly matching the underlying model, yielding ARI = 1. For SD = 0.2 $\times$ mean (Fig. \ref{fivestateSDsstudy}E), the five-cluster solution retains near-complete recovery with broader confidence ellipses (ARI = 0.95), while the two-cluster solution merges multiple populations (ARI = 0.2, result not shown). For SD = 0.3 $\times$ mean (Fig. \ref{fivestateSDsstudy}F), only two clusters are recovered (ARI = 0.2), with cluster centers at (500 cps, 0.8 ns) and (1920 cps, 1.65 ns), demonstrating substantial merging. These results illustrate that as states become more diffuse (characterized by larger SDs), state overlap increases, which reduces ground-truth recoverability, defining the practical breaking point of the clustering protocol, i.e., when states begin to overlap significantly with one another, clustering becomes more ambiguous.}\\

{Before proceeding to explore the effect of varying state probabilities, we tested the robustness of alternative clustering methods on the well-defined system shown in Figure \ref{fivestateSDsstudy}D, where GMM with IC metrics perfectly recovered all five populations. For the DPGMM comparison, the weight concentration prior was again set to 0.1. When the number of prior components was set to six, all five clusters were recovered correctly, as shown in Figure ~\ref{compareclustermethodtwo}A. However, when the number of priors was increased to eight, mimicking a situation in which the true number of clusters may be unknown, the model selected seven clusters (Fig.~\ref{compareclustermethodtwo}B), assigning two low-occupancy “ghost” clusters in addition to the true populations, highlighting its sensitivity to prior specification. For K-means clustering, the elbow plot is shown in Figure ~\ref{compareclustermethodtwo}C, with gradient lines used to identify the elbow. Based on this gradient method, three or four clusters appear as plausible solutions, but the clustering results for these choices (Figs. ~\ref{compareclustermethodtwo}D and E) are clearly poor, failing to recover the ground-truth populations. When five clusters are explicitly specified (Fig.~\ref{compareclustermethodtwo}F), K-means successfully recovers all populations; however, this requires prior knowledge of the correct number of clusters, which is undesirable for an objective clustering framework. Together, these comparisons underscore both the sensitivity of DPGMM to prior settings and the limitations of K-means in objectively selecting the correct number of clusters, even when the underlying populations are well-defined.}\\

\noindent {\textbf{Five-state system: effect of relative state probabilities}}\\

\begin{figure}[h!tbp]
    \centering
    \includegraphics[width=1\linewidth]{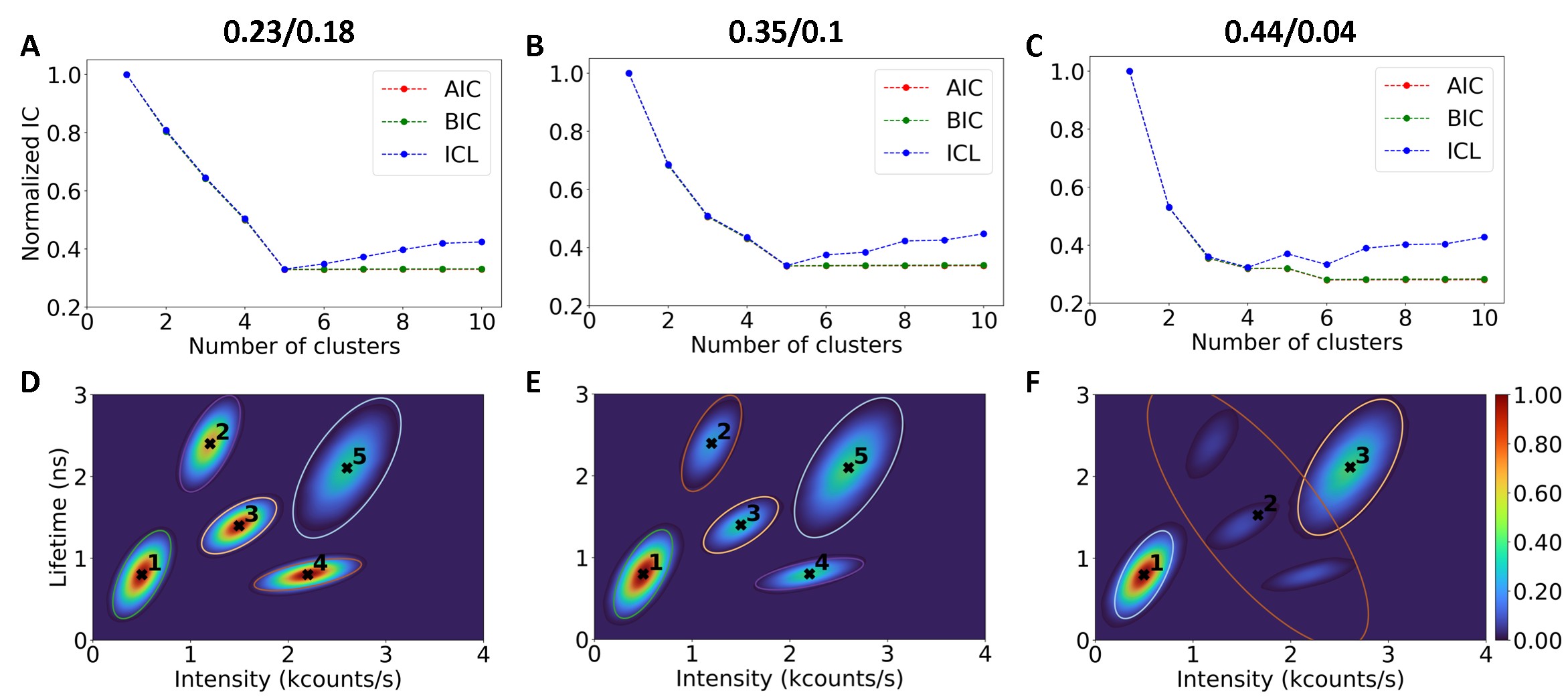}
    \caption{{Effect of changing the relative probabilities of the two initial, high-occupancy states versus the three additional, low-occupancy states on ground-truth recovery. A--C. IC score plots for two initial-state/additional-state probability combinations, where the first value refers to the probability of the two initial states (1 and 5) and the second to the three additional states (2--4): 0.23/0.18 (A), 0.35/0.1 (B), and 0.44/0.04 (C). For the left and middle cases (A, B), the IC curves consistently identify the five-cluster solution as optimal, supported by near-ideal tightness ($\langle D^2\rangle \approx 2$) and FO $\approx 0.05$. In contrast, for the extreme occupancy imbalance in (C), a three-cluster solution is selected as optimal ($\langle D^2\rangle \approx 1.97$, FO $\approx 0.04$). D--F. Corresponding clustering results: in (D) and (E), all five ground-truth states are perfectly recovered (ARI = 1), with stable cluster centers (black crosses) and progressively reduced population of the three additional states. In (F), the two initial states are still recovered accurately, while the three low-occupancy, diffuse states are merged into a single cluster, reflecting effective down-weighting under strong probability imbalance.}}

    \label{fivestateProbstudy}
\end{figure}

\noindent {To investigate the influence of relative state occupancy on cluster recoverability, we used the system described in Figure \ref{fivestateSDsstudy}D, with SD = 0.1 $\times$ mean and where all states have a probability of occurrence set to 0.2, as a baseline, and systematically varied the relative probabilities of the two initial states (1 and 5) with respect to those of the three additional states (2--4). The results are summarized in Figure \ref{fivestateProbstudy} with initial/additional state probability combinations of 0.23/0.18 (left column), 0.35/0.1 (middle column), and 0.44/0.04 (right column). In these simulations, the positions and widths of all states remained fixed, so the degree of overlap was unchanged; only the relative state probabilities of the two initial and three additional states were altered. Figure \ref{fivestateProbstudy} A--C correspond to the IC score plots for the respective cases. For the left and middle probability configurations, the IC scores consistently identify the five-cluster solution as the preferred choice, supported by near-ideal cluster tightness ($\langle D^2\rangle \approx 2$) and FO $\approx 0.05$, indicating that moderate changes in occupancy do not affect either the preferred cluster number or the intrinsic solution quality. The corresponding clustering results (Figs.~\ref{fivestateProbstudy}D and E) show perfect recovery of all five ground-truth states, with the two initial states and the three additional states remaining clearly resolved despite their differing occupancies. In contrast, for the extreme probability imbalance in the right column (0.44/0.04), the IC analysis no longer identifies the five-cluster solution as a viable candidate. Instead, a three-cluster solution is selected as optimal, with $\langle D^2\rangle \approx 1.97$ and FO $\approx 0.04$, reflecting adequate compactness and data coverage. A two-cluster solution also appears as a candidate, exhibiting slightly stronger compactness ($\langle D^2\rangle \approx 2$) but a substantially elevated FO ($\approx 0.09$), indicating that a significant fraction of points fall outside the confidence regions, an undesirable outcome relative to the target FO $\approx 0.05$ and indicative of under-partitioning. The clustering result corresponding to the selected three-cluster solution is shown in Fig.~\ref{fivestateProbstudy}F, where the two initial (well-defined) states are recovered cleanly, while the three low-occupancy states are grouped into a single composite cluster. Together, these results demonstrate that while state probabilities generally influence the statistical prominence of individual populations, clustering performance is largely insensitive to occupancy changes when geometric overlap is fixed. Only when a pronounced density imbalance is introduced, due to relative probability differences, are low-occupancy states effectively down-weighted, leading to their controlled aggregation without compromising the recovery of the dominant, well-defined states.}
\\

\noindent {\textbf{Final test: effect of blinking models on a complex dataset}}\\

{To assess whether blinking kinetics influence clustering performance in highly congested datasets, we generated an overlap-dominated ten-state system using different on-state dwell-time models. The dataset combines five well-defined states with five diffuse states, all assigned equal probabilities, with narrow and comparable SDs for the well-defined states and broader distributions for the additional states; polarization effects were included. Mean values and SDs are listed in Table \ref{noiseltdsimtable}, with states 1--5 corresponding to well-defined populations and states 6--10 to diffuse states. Figure \ref{blinkingmultistate}A--C shows the results for power-law on- and off-state blinking, while panels D--F, G--I, and J--L correspond to exponential, truncated power-law, and high-exponent truncated power-law on-state models, respectively, with off-state dwell times following power-law statistics in all cases. Across all blinking models, the clustering outcome is effectively unchanged. In all cases, three clusters are recovered at similar centers (at approximately 510, 1170, and 2020 cps with corresponding lifetimes of 0.71, 2.3, and 1.45 ns, respectively), with mean cluster tightness in the range $\langle D^2\rangle \approx 1.80$–1.87, FO $\approx 0.03$, and ARI values of 0.11–0.12. The failure to recover the ground-truth states, therefore, does not originate from the choice of blinking kinetics but reflects an intrinsic limitation imposed by strong population overlap and distribution broadening. In this regime, clustering performance is governed by the geometric structure of the lifetime–intensity distributions rather than by the underlying temporal blinking statistics.}\\

\noindent {\textbf{Broad applicability}}\\

{Our clustering protocol is broadly applicable to diverse single-molecule observables beyond lifetime--intensity data, including any measurement where populations can be represented as points in a multidimensional space and exhibit spread or overlap. To demonstrate this, we simulated a spectral dataset based on peak position and full-width at half-maximum (FWHM) values for multiple states, sampling each within designated noise levels to mimic experimental variability. Figure \ref{fwhmSI}A shows a scatter plot of FWHM--peak correlations, emulating a possible experimental result, while Figure \ref{fwhmSI}B presents the corresponding two-dimensional ensemble-averaged spectrum. Figure \ref{fwhmSI}C displays a density plot of the FWHM--peak position distribution. When applied to this dataset, the clustering protocol accurately identifies the statistically optimal number of populations. The IC score plots in Figure \ref{fwhmSI}D consistently select four clusters, with $\langle D^2\rangle = 2$ and FO = 0.06. The resulting cluster assignment, shown in Figure \ref{fwhmSI}E, recovers all underlying states with ARI = 1, confirming high fidelity even under realistic noise conditions.}

\subsection*{Alexa}
\begin{figure}[h!tbp]
    \centering
    \includegraphics[width=0.7\linewidth]{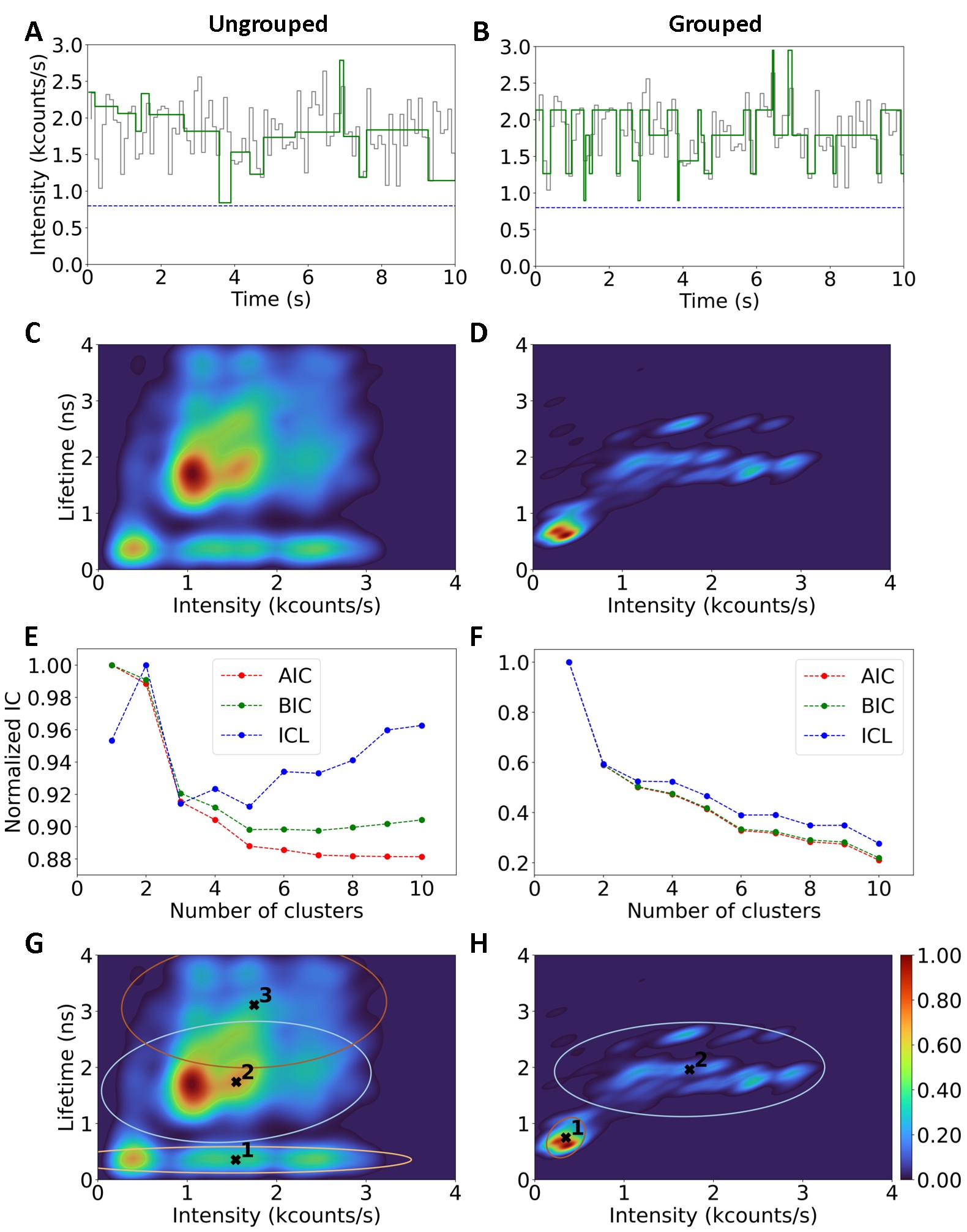}
    \caption{Comparison of the performance of the clustering protocol on the ungrouped (left column) and grouped (right column) Alexa data with 52 particles. A and B. Example intensity traces over 10-s windows, with grey representing the 40-ms binned intensity data, {green showing the ungrouped (A) or grouped (B) trace, and the dashed blue line indicating the background intensity level}. C and D. Lifetime--intensity distributions for the corresponding datasets. E and F. {IC score plots used to select the optimal number of clusters; for the ungrouped data, three clusters are preferred, whereas for the grouped data, two clusters are chosen. Clustering outcomes for the selected cluster numbers, with cluster centers indicated by black crosses and 0.9 confidence ellipses showing the spread of data points assigned to each cluster. }}
    \label{AlexaSixPanel}
\end{figure}

We applied our clustering protocol to both the ungrouped and grouped-level Alexa data to demonstrate the effectiveness of the grouping technique described in Ref. \cite{botha2024advanced}. The Alexa dataset was chosen as a case study of rapid photoblinking, with dwell times in intensity states generally too short to be resolved visually or with a statistical change-point analysis (Fig. \ref{AlexaSixPanel}A). This was attained by deliberately not removing oxygen from the buffer solution. In addition, spin-coating of the sample introduced a random distribution of dipole orientations. We used circularly polarized excitation light to consider only the polar angle dependence of the dipole orientation, similarly to Figure \ref{smearedS1}. The resulting erratic intensity behavior and data smearing are shown in Figure \ref{AlexaSixPanel}, with the left column representing the resolved-level (ungrouped) data and the right column representing the grouped-level data. Clearly, change-point analysis of the data had difficulty resolving the actual intensity levels accurately (Fig. \ref{AlexaSixPanel}A, green trace), whereas this improved markedly after grouping the data using the built-in grouping feature of \textit{Full SMS} (Fig. \ref{AlexaSixPanel}B, green trace). The lifetime--intensity distribution of the ungrouped data (Fig. \ref{AlexaSixPanel}C) was also significantly more spread out than that of the grouped data (Fig. \ref{AlexaSixPanel}D) and showed little discernible trend, making it difficult to distinguish between individual populations. In contrast, the unbiased grouping (Fig. \ref{AlexaSixPanel}D) statistically assigns similar resolved levels to a single grouped level, which effectively denoises the data and facilitates easier and more accurate clustering. Furthermore, the grouping also removed most of the short-lifetime data in the ungrouped distribution, which highlights the large fitting uncertainties when dealing with low photon budgets.

{For the resolved ungrouped Alexa data, all three IC scores (Fig. \ref{AlexaSixPanel}E) indicate that three and five clusters are the most statistically prominent local minima. Considering the corresponding auxiliary metrics of $\langle D^2\rangle = 1.81$, FO = 0.07, and $\langle D^2\rangle =1.71$, FO = 0.06, respectively, for a confidence interval of 0.9, the three-cluster solution is optimal. The resulting three-cluster analysis of the ungrouped data (Fig. \ref{AlexaSixPanel}G) reproduces the expected broad structure, with cluster centers corresponding to the mean values within the confidence ellipses rather than the densest populations, due to the inherent spread of the data.} {For the grouped data, the IC score plots decrease with increasing cluster number. Using a standard IC-based approach that selects the global minimum as the optimal solution tends to favor higher cluster numbers, which would clearly lead to overfitting in this case. Our algorithm identifies two clusters as the optimal solution, since it corresponds with $\langle D^2\rangle =2$ and FO = 0.09, while all other candidate solutions, including three, six, eight, and ten clusters, yield lower tightness values ($\langle D^2\rangle< 1.94$) and smaller fractions outside (FO~$<0.08$), suggesting over-segmentation and reduced interpretability. The result of using two clusters for the grouped data is shown in Figure \ref{AlexaSixPanel}H, which produced more coherent and visually consistent cluster centers than the three-cluster solution in Figure \ref{AlexaSixPanel}G  while avoiding overfitting.}

To eliminate extreme outliers, we implemented the modified IQR outlier detection rule (described in Eq. \eqref{IQRrule}) to identify data points that lay unusually far from their respective cluster centers. This procedure eliminated roughly $2\%$ of the data points, resulting in a retention of {$\sim97\%$}, and ensured that subsequent calculations of any cluster-based statistics were not skewed by outliers, while keeping the cluster assignments unchanged. For the grouped data, clustering revealed two states with lifetimes of {$0.75\pm0.17 \, \text{ns}$ and $1.94\pm0.35 \, \text{ns}$}, respectively, with an average weighted lifetime of $1.73\pm0.55 \, \text{ns}$. There are $\frac{N(N-1)}{2}$ possible reversible pairwise switching pathways for $N$ clusters. {The switching rates between the two states were $k_{12} = 9.87 \,\text{s}^{-1}$ and $k_{21}  = 1.99\,\text{s}^{-1}$, with a switching frequency of $f = 3.31\,\text{s}^{-1}$, and switching ratio of $R_{12}>>1$ indicating that switching from state 1 to state 2 is more likely, i.e., state 2 is the more stable state, in agreement with its higher population density.}

\subsection*{QD 605}
\begin{figure}[h!tbp]
    \centering
    \includegraphics[width=0.7\linewidth]{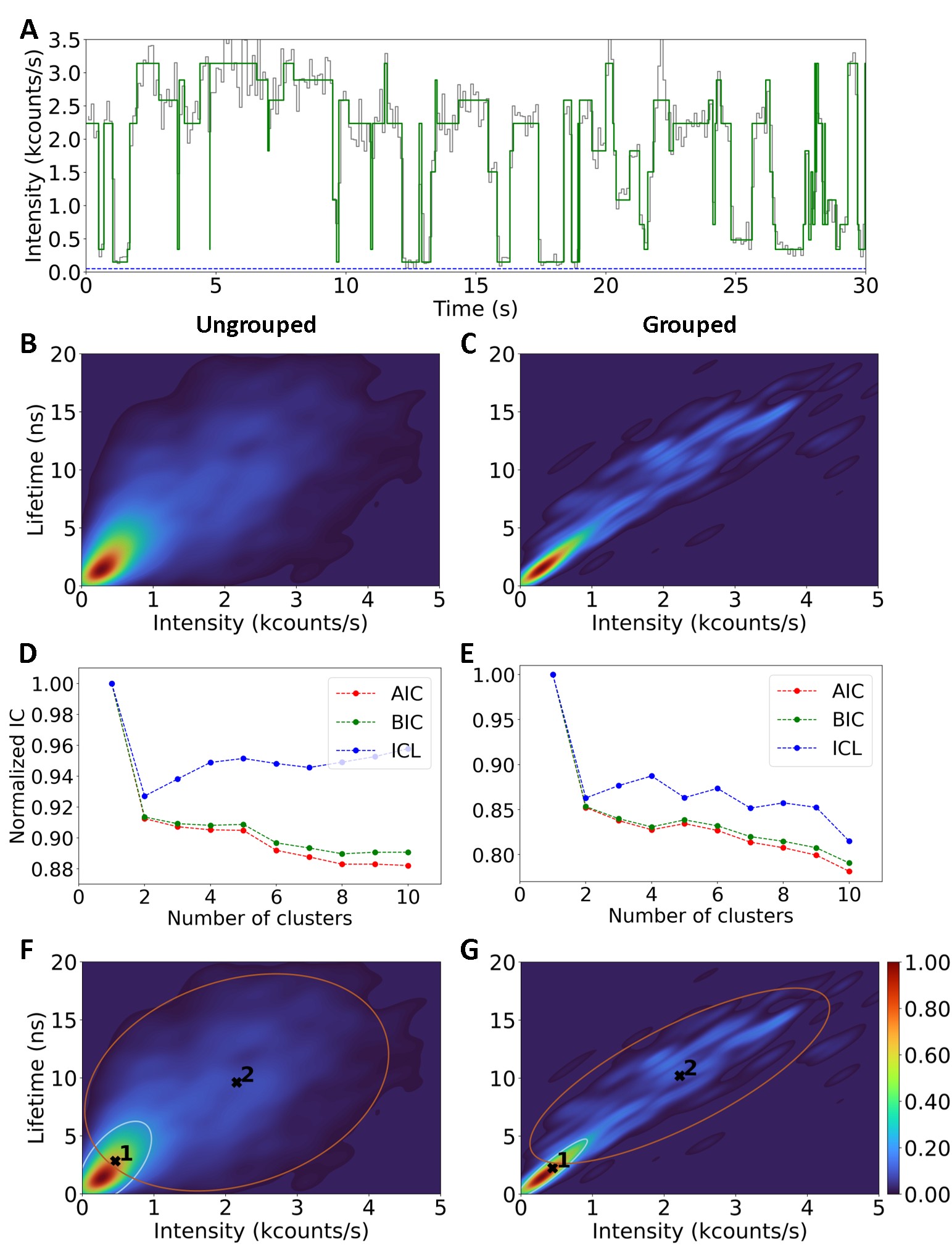}
    \caption{Clustering results for the statistically small QD 605 data with $\sim 30$ particles. A. Representative example intensity trace showing 40-ms binned data (grey) and resolved intensity levels (green), {with the dashed blue line indicating the background intensity level}. B and C. Lifetime--intensity distribution for the ungrouped (B) and grouped (C), respectively. D and E. {Corresponding IC score plots.} F and G. Clustering both the ungrouped and grouped datasets using two clusters, with cluster centers indicated by black crosses, and ellipses (at a confidence of $0.85)$ indicating the assignment of data points to each cluster.}
    \label{QdotGroupclustering}
\end{figure}

To test the robustness of the clustering analysis to statistically small datasets, we considered a dataset comprising only 30 QDs, measured for 30 s each. QD 605 consists of a CdSe core surrounded by a ZnS shell. When the shell is sufficiently thin, these QDs, which belong to type II-IV QDs, typically exhibit bimodal blinking behavior \cite{palstra2021python, efros2016origin}. A representative intensity trace is shown in Figure \ref{QdotGroupclustering}A, illustrating the relatively fast photoblinking dynamics of this sample, although, unlike the Alexa data, the intensity levels could be well-resolved by the change-point algorithm. The lifetime--intensity distributions for the ungrouped and grouped QD data are shown in Figure \ref{QdotGroupclustering}B and C, respectively. The grouping effectively reduces the noise in the data, as evidenced by the markedly narrower lifetime--intensity distribution, which uncovers the underlying linear correlation between the two parameters. The significant denoising facilitated the subsequent data clustering. {The IC score plots for the ungrouped and grouped data are shown in Figures \ref{QdotGroupclustering}D and E, respectively. For this dataset, it is again evident that a global minimum solution, which lies even beyond ten clusters, will yield over-partitioning; hence, we instead employ our candidate cluster identification method. For the ungrouped data, the BIC and AIC curves show a shallow gradient beyond two clusters, while ICL deviates, rising after two clusters. Universally, the first steepest drop (with the greatest IC gain) occurs at two clusters, yielding $\langle D^2\rangle =1.89$ and FO = 0.13, indicating that two clusters are the most statistically significant choice. While AIC and BIC suggest eight clusters as a secondary option, the auxiliary metrics of $\langle D^2\rangle =1.6$ and FO = 0.08 deviate too much from their optimal values, indicating that the two-cluster solution provides the best optimal choice between compactness and coverage of the data, for the candidate numbers of clusters. For the grouped dataset, AIC and BIC again show a near-identical trend with the strongest elbow at two clusters, while ICL suggests additional possible partitions at five, seven, and ten clusters. However, the two-cluster solution (with $\langle D^2\rangle =1.89$ and FO = 0.11) remains the most meaningful partition of the grouped data, representing physically relevant states while avoiding overfitting. Figures \ref{QdotGroupclustering}F and G show the resulting two-cluster solutions for the ungrouped and grouped data, respectively.}

We implemented the modified IQR outlier detection rule (Eq. \eqref{IQRrule}) on the grouped data prior to calculating the switching rates and frequencies. Approximately $4\%$ of the data points were removed, resulting in a retention of $\sim96\%$ of the original data. This step ensured that extreme outliers were disregarded so that they would not skew the computed dynamics. The lifetimes corresponding to the two cluster centers of the grouped data are $2.2\pm1.2 \, \text{ns}$ and $9.5\pm3.3 \, \text{ns}$, with a weighted average of $7.0\pm4.6 \, \text{ns}$. The switching rates between the two states were $k_{12} = 1.68 \,\text{s}^{-1}$ and $k_{21}  = 3.04\,\text{s}^{-1}$, with a switching frequency of $f = 2.16\,\text{s}^{-1}$, {and with a switching ratio of $R_{12}<1$} indicating that switching from state 2 to state 1 is more likely, i.e., state 1 is the more stable state, in agreement with its higher population density.

\section{Discussion}

Owing to the high sensitivity of SMS, {even modest experimental noise, system heterogeneity, suboptimal experimental conditions, or fitting uncertainties can substantially broaden measured data distributions.} This is especially true for fluorescence intensity data, as it is generally strongly dependent on the excitation probability of the sample \cite{kruger2011fluorescence}. On the other hand, the resolved lifetime usually has a greater uncertainty than the resolved intensity, leading to additional broadening {along the lifetime axis. Consequently, such broadening is most often dominated by noise and measurement or analysis uncertainty rather than genuine heterogeneity in the underlying physical states.}

Skewed or unevenly distributed populations can bias clustering, causing cluster centers to be shifted away from where they might be expected through visual inspection. To address this, the grouping algorithm, implemented in the \textit{Full SMS} software, helps to denoise the data and reveal any dominant underlying relationships between experimental parameters, thereby tightening distributions for interpretation. The effect is evident when comparing, for example, Figures \ref{AlexaSixPanel}C and \ref{AlexaSixPanel}D, where, in the latter case, a cleaner and more structured distribution is observed. Grouping also emphasizes the linear trend between lifetime and intensity in many SMS experiments \cite{schlau2015single, botha2024advanced, gwizdala2016controlling, weston2002measuring, tinnefeld2001photophysical}. By denoising and tightening the data, grouping makes subsequent clustering easier, more robust, and more reliable. 

{Comparison of alternative clustering methods on well-defined simulated datasets (Figs. \ref{compareclustermethodone} and \ref{compareclustermethodtwo}) highlights the advantages of the GMM-based protocol. SMS data are naturally elliptical with non-zero covariance between intensity and lifetime, a structure that GMMs inherently capture through full covariance estimation, whereas DPGMM and K-means either depend sensitively on priors or assume spherical clusters. DPGMM can recover the underlying populations but is highly sensitive to prior specification, which can lead to low-occupancy “ghost” clusters as observed in Figures \ref{compareclustermethodone}A and \ref{compareclustermethodtwo}B. K-means, being a hard-assignment method, relies on subjective elbow-point selection and does not provide probabilistic assignments, making it prone to misclassification unless the number of clusters is manually set (see Fig. \ref{compareclustermethodtwo}). Both methods are also computationally slower than GMM with IC-based selection. Overall, GMM with IC metrics provides more reliable, objective, and efficient clustering of SMS datasets without requiring prior knowledge of the number of populations.}

It is important to note that the clustering algorithm does not assess whether the clusters correspond to physically meaningful states or not; it simply partitions the data based on localized high-density regions. {The protocol models the data as a Gaussian mixture, with the optimal cluster number determined using multiple IC metrics (AIC, BIC, and ICL), which balance goodness-of-fit, model complexity, and classification uncertainty. For SMS datasets in general, primarily due to their intrinsic noise and measurement uncertainty, IC score plots often exhibit several candidate elbow points rather than a single monotonic trend. While the global minimum provides the statistically best fit, in experimental datasets, it may correspond to unrealistically high cluster numbers that result in over-partitioning and lack physical interpretability (see, for example, Figs. \ref{AlexaSixPanel}E--F and  \ref{QdotGroupclustering}D--E). The clustering algorithm removes a substantial portion of subjectivity by automatically determining optimal candidates based on relevant IC gain and steep gradient descent in the plots. To further mitigate ambiguity, auxiliary cluster-quality metrics---average cluster tightness and FO---are evaluated alongside the IC scores. Cluster tightness quantifies compactness relative to dimensionality, while FO assesses whether the confidence regions adequately capture the data spread. This combined framework ensures that selected solutions are statistically consistent and interpretable, even when IC trends alone do not uniquely identify an optimum. In the Alexa dataset (Figs.~\ref{AlexaSixPanel}E and F), multiple candidate solutions are present; however, only the selected solutions simultaneously maintain reasonable tightness and data coverage, preventing over-segmentation and yielding physically interpretable results. While AIC and BIC generally exhibit similar trends, ICL often diverges beyond the most prominent elbow points. This suggests that, for practical purposes, AIC and/or BIC alone may be sufficient in future studies.}

\subsection*{Simulations}

{The simulated benchmarks were constructed as a sequence of controlled stress tests, progressing from idealized, well-separated systems to increasingly overlap-dominated and experimentally distorted regimes, in order to identify both the operational range and the intrinsic limits of the clustering protocol. Because the ground truth is known in all simulations, these tests allow direct distinction between algorithmic performance and fundamental limits imposed by data geometry. For clean two- and four-state systems with equal occupancies and minimal overlap (Figs.~\ref{twostatesim} and \ref{fourstatesim}), the protocol consistently recovers the correct number of clusters, with ideal compactness ($\langle D^2\rangle = 2$), FO values near the expected 0.05, and unity ARI. These results establish a baseline: when populations are statistically well-separable, the IC metrics, confidence ellipses, and auxiliary quality measures are mutually consistent, and recovery is unambiguous. Importantly, increasing the number of well-defined states does not degrade performance, confirming that the protocol scales correctly when overlap remains low. Polarization and dipole-orientation effects introduce the first nontrivial distortion by broadening intensity distributions without adding new physical states (Fig.~\ref{smearedS1}). Importantly, an additional cluster may arise that does not correspond to a true physical state but arises from the asymmetric intensity distribution created by polarization effects. This demonstrates that immediate, polarization-based broadening contributes to apparent cluster splitting, and as shown in subsequent multi-state simulations, such effects can compound with diffuse populations and overlap to produce unexpected or “nonphysical” clustering behavior.}

{The influence of geometric overlap was isolated by adding additional states at fixed occupancy while systematically varying only their distribution widths (Fig.~\ref{fivestateSDsstudy}). These results indicate that as the state overlap increases, clustering progressively becomes more ambiguous due to reduced separability between underlying states. Importantly, the clustering method remains non-restrictive: higher-cluster solutions (e.g., five clusters) remain accessible if desired, but are automatically flagged as statistically weak, thereby avoiding subjective selection while preserving user flexibility. The loss of ground-truth recoverability at SD = $0.3\times$ mean is also physically expected, as such extreme broadening is unlikely under typical experimental conditions. However, significant overlap may occur in experimental datasets comprising a larger number of states with more moderate broadening, although experimentally induced broadening can often be mitigated through grouping strategies, as demonstrated in Fig.~\ref{AlexaSixPanel}D, which effectively tightens populations.}

{Varying state occupancies at fixed distribution widths primarily alters relative population density rather than geometric separability, as observed in Fig.~\ref{fivestateProbstudy}. For small to moderate occupancy variations, the preferred cluster number and the intrinsic quality of candidate solutions---quantified by cluster tightness and coverage---remain essentially unchanged, indicating that clustering performance is largely insensitive to modest probability shifts when overlap is fixed. However, for sufficiently large occupancy imbalances, a qualitative change emerges: low-probability states are effectively down-weighted and are no longer resolved individually, but instead become coupled into a single cluster, while the high-occupancy, well-defined states remain accurately recovered. Unlike distribution broadening, which directly increases overlap and limits recoverability across all states, changes in occupancy influence clustering only once extreme probability disparities are reached. Overall, the absolute probability of occurrence has a far weaker impact on clustering performance than the distribution width and the resulting geometric overlap.}

{Across all simulation classes, including those with alternative blinking kinetics, compactness and FO remain stable while ARI degrades only when overlap becomes dominant, confirming that clustering performance is controlled by the geometric structure of the observable space rather than by dwell-time statistics. Furthermore, benchmarking across multiple blinking models confirms that variations in on- and off-state dwell-time distributions have minimal impact on the inferred clustering, as the algorithm is governed primarily by the geometric structure of the lifetime–-intensity distributions rather than by the detailed temporal kinetics of state switching. Together, these results demonstrate that the proposed clustering framework is both robust and broadly applicable under diverse experimental conditions.}

\subsection*{Alexa }
For Alexa, the ungrouped data (Fig. \ref{AlexaSixPanel}C) appear more spread out with no obvious trend, whereas the grouped data (Fig. \ref{AlexaSixPanel}D) reveals the underlying trends due to substantial denoising. The large intensity spread in the grouped data corresponding to a lifetime of $\sim1.8-2.0$ ns can be explained by Figure \ref{smearedS1}, i.e., a strong broadening in the intensity distribution of the bright state is caused by the heterogeneous distribution of molecular orientations relative to the circularly polarized excitation light, causing large variations in the relative absorption cross-sections of the molecules. The absence of broadening down to the background level can be explained by user bias in the data selection: molecules that are oriented away from the focal plane interact only weakly with the incident light and, therefore, appear to be significantly dimmer than those oriented near-parallel to the focal plane. These dim signals are easily masked by the background noise and, as a result, those molecules are unlikely to be selected for further study. This skews the overall intensity distribution to brighter states, as is the case for Alexa (Fig. \ref{AlexaSixPanel}D).

The authors in Ref. \cite{wang2013lifetime} report the lifetime of Alexa 647 to be $1.13 \, \text{ns}$, which is longer than the bulk lifetime of 1 ns due to the use of glycerol. In our study, the sample was spin-coated and measured in the presence of oxygen. The average weighted SM lifetime in our dataset ($1.73\pm0.55 \, \text{ns}$) deviates from that of previous studies, although this prolonged lifetime is not likely caused by the effects of different environmental conditions, but instead a result of the data grouping by the \textit{Full SMS} software. Figures \ref{AlexaSixPanel}C and D indicate that the average lifetime of the quenched state (state 1) increased from $\sim 0.35$ ns for the ungrouped data (Fig. \ref{AlexaSixPanel}C) to $\sim 0.75$ ns for the grouped data  (Fig. \ref{AlexaSixPanel}D). This indicates that most of the short-lifetime data in the ungrouped dataset was combined with longer-lifetime data during the grouping procedure, likely to limit the number of short events that lack statistical weight or are prone to fitting errors. This unavoidable artifact of data grouping can be limited by choosing another IC value during the grouping analysis {(in the \textit{Full SMS} software)} to ensure that a short-lifetime population is chosen as a distinct state. 

{State 2 represents a single physically broadened state}; however, previous studies \cite{lin2015quantifying, yang2024electrochemically, herdly2023benchmarking} suggested that Alexa may exhibit multiple distinct fluorescing states. {These additional states may be masked by the broadened population, emphasizing the importance of ensuring optimal experimental conditions. Nevertheless, the presence of multiple candidate solutions in the IC scores demonstrates that our clustering protocol can reliably identify and separate physically meaningful subpopulations, thereby allowing the user to select an alternative elbow point if prior knowledge about the expected number of states is available.}

\subsection*{QD 605}
We used QDs as a benchmarking sample, as they are widely used in photoblinking studies. In general, the blinking behavior of QDs is heavily influenced by factors such as their core composition \cite{palstra2021python}, shell thickness \cite{efros2016origin}, and the excitation power. These parameters influence the number and stability of accessible quasi-stable lifetime--intensity states. QD 605, used for the current study, has a CdSe core tuned to emit around 605 nm and capped with a thin ZnS shell with a thickness of $<2$ nm. Light absorption arises from excitonic transitions in the CdSe core, with multiple higher-energy transitions giving rise to significant broadening of the absorption toward the blue. The ZnS shell is transparent in this range and primarily serves to passivate any surface states and confine carriers. QDs with relatively thin shells generally follow Type-A blinking behavior \cite{efros2016origin}, which is driven by Auger recombination, i.e., recombination of an electron-hole pair that involves transfer of the energy to a different charge carrier instead of photon emission. This Type-A blinking is, therefore, associated with blinking from a lower-intensity, short-lifetime state (state 1) to a higher-intensity, longer-lifetime state (state 2), and hence the two states show a linear relation between the lifetime and intensity, in line with Figure \ref{QdotGroupclustering}C. The high density of the Auger-dominated state 1 arises because these datapoints occupy a narrow region in lifetime--intensity space, making the population appear tighter. The results of the clustering protocol applied to the QD data indicated that two clusters were optimal (see Fig. \ref{QdotGroupclustering}E), and the weighted single-molecule average lifetime was found to be $7.3 \, \pm4.6 \,\text{ns}$. Ref. \cite{grecco2004ensemble} reports QD 605 lifetimes to be in the range of $10-20$ ns depending on excitation power and other experimental conditions, while the bulk (solution-phase) lifetime was measured as $6.3\pm 1.2 \, \text{ns}$ in Ref. \cite{bhuckory2016evaluating}. Our experimental conditions were similar to those in the latter study, with QDs dispersed, no oxygen scavengers or additional additives such as glycerol used. Therefore, our measured average lifetime is in good agreement with the bulk value. The relatively high uncertainty in our lifetime arises primarily from the statistically small size of the dataset and the heterogeneity of the SM events. To fully establish the exact number of relevant states for this QD sample, one would have to systematically compare the switching dynamics and blinking behavior for different choices of the cluster number. Nevertheless, the two-cluster fitting is sufficient based on the switching dynamics (switching rates and frequencies) between these two clusters, which indicates a strong reversible blinking mechanism.

\section{Conclusions}

While highly sensitive, SMS is prone to significant noise, which can easily introduce fitting errors, bias in the parameter estimates, and artificially inflated variability. This reduces the ability to identify true subpopulations, leading to higher uncertainty and weaker correlations. It is, therefore, crucial to perform careful and reliable noise reduction to preserve meaningful structure while minimizing the spurious effects of noise, without introducing new biases. {Due to the specific structure of SMS data, our GMM and IC-based clustering protocol performs significantly better than K-means and DPGMM (a recommendable Bayesian method) in objectively identifying the statistically most likely number of clusters in noisy SMS datasets. We have benchmarked our protocol across a range of simulated and experimental datasets to validate its performance under diverse conditions.}

{For clean two- and four-state systems, the protocol reliably recovers the correct number of clusters and underlying state distributions, where cluster-quality metrics and agreement with ground truth are simultaneously optimized. Polarization and dipole-orientation effects introduce asymmetry and density shifts, producing apparent broadening without adding physical states; despite this, the protocol continues to identify statistically stable, well-supported partitions that reflect the dominant structure of the data. Recovery is robust for moderate state overlap (APGO$<0.2$), but at high overlap (APGO$>0.4$), individual states become statistically inseparable, leading the protocol to select fewer, broader clusters. Varying relative occupancies at fixed widths has a minimal effect on clustering unless extreme imbalances down-weight low-probability states, which are then merged while well-defined states remain accurately recovered. These results show that distribution width and geometric overlap are the primary determinants of clustering efficacy, with occupancy effects relevant only in extreme cases.}

{Across all simulation classes, including alternative blinking kinetics, clustering outcomes are largely insensitive to dwell-time distributions, underscoring the robustness of the protocol for diverse single-molecule datasets. In experimental datasets, where ground truth is inherently inaccessible, cluster selection must rely on statistically grounded criteria. Solutions exhibiting expected cluster tightness and appropriate data coverage provide the most defensible interpretation, even when alternative partitions exist. This emphasizes that statistical optimality is the appropriate objective for experimental single-molecule data. For Alexa 647, three clusters were identified in the ungrouped data, which reduced to two clusters after grouping, demonstrating that grouping improves denoising and suppresses fragmentation of noisy or sparsely populated states. For QD 605, two clusters were consistently recovered across simulations and experimental data, confirming stable detection of physically meaningful subpopulations. }

{Across all datasets, AIC and BIC generally exhibit consistent trends and minima, while ICL often follows a slightly different trajectory due to its stronger penalty for overlap; however, shared elbow points indicate that AIC or BIC alone may suffice for practical cluster selection. The protocol also facilitates identification and removal of misfit clusters, clusters arising from nonoptimal sample preparation, and minor clusters obscured by noise. Non-monotonic behavior of BIC and related IC scores further emphasizes that multiple local minima are expected; cluster selection should therefore be guided by a combination of IC trends and cluster-quality metrics rather than by a single global minimum. Follow-up studies exploring how different statistically defined clusters influence inferred switching dynamics would be valuable.}

{In this context, the goal of clustering is not the exhaustive reconstruction of every hypothetical underlying physical state, but the identification of statistically defensible, experimentally interpretable subpopulations supported by the available information. The proposed protocol is explicitly designed to operate within this constraint: it does not enforce a fixed cluster number, nor does it privilege formal global minima of information criteria when these correspond to over-partitioned or physically ambiguous solutions. Instead, it identifies a set of statistically plausible candidate solutions and evaluates them using independent measures of compactness and data coverage, thereby distinguishing genuine structure from artifacts of noise, overlap, or density imbalance. This distinction is particularly important for experimental SMS datasets, where the true number of states is fundamentally unknowable, and overfitting can easily masquerade as increased physical insight. By prioritizing statistical identifiability over nominal state count, the protocol provides a principled and minimally subjective framework for clustering noisy, heterogeneous single-molecule data. When the data support fine structure, it is recovered; when they do not, the method reports this limitation transparently rather than producing spurious resolution, especially when compared to other clustering methods. As such, the framework offers a robust foundation for downstream analyses of switching dynamics, state interconversion, and photophysical mechanisms across a broad range of single-molecule experiments.}

\section{Supporting Material}

The Supplementary Information contains six supportive figures and one table.

\section{Acknowledgments}
MACL was supported by the National Research Foundation (NRF), South Africa (PMDS22070633262). JLB was supported by the Vrije Universiteit Amsterdam-NRF Desmond Tutu programme (grant no. 99413). BvH was supported by the NRF (grant nos. 115463 and 120387). TPJK acknowledges funding from the NRF (grant nos. 137973 and 0403211945) and the Rental Pool Programme of the Council for Scientific and Industrial Research’s Photonics Centre, South Africa. 

\section{Author Contributions}

All authors conceived the study. BvH performed the experiments and captured the data. JLB performed pre-clustering analyses. MACL developed the clustering software, performed the simulations, and analyzed the data. TPJK acquired funding, supervised research, and administered the project. MACL and TPJK wrote the manuscript. All authors read and approved the final manuscript.

\section{Declaration of interests}

The authors declare no competing interests.

% Uncomment if using bibtex (default)

% Uncomment if using biblatex
% \printbibliography

%\documentclass{article}
%\usepackage[utf8]{inputenc}
%\usepackage[colorlinks,allcolors=cyan!70!black]{hyperref}
%\usepackage[normalem]{ulem}
%\usepackage{xcolor}
%\usepackage{authblk}

%\renewcommand{\thefootnote}{\fnsymbol{footnote}}
%\newcommand{\gtwo}{$g^{(2)}(\tau) $}

% https://universitypretoria-my.sharepoint.com/:w:/g/personal/u13290152_up_ac_za/EZgmGojyEH9GhzBc0WnchcoBRa4RaqgLnBKQcDM_CI26Jw

%\usepackage[top=15mm,bottom=30mm,left=30mm,right=30mm]{geometry}
%\usepackage{graphicx}%
%\usepackage{multirow}%
%\usepackage{amsmath,amssymb,amsfonts}%
%\usepackage{amsthm}%
%\usepackage{mathrsfs}%
%\usepackage[title]{appendix}%
%\usepackage{xcolor}%
%\usepackage{textcomp}%
%\usepackage{manyfoot}%
%\usepackage{booktabs}%
%\usepackage{algorithm}%
%\usepackage{algorithmicx}%
%\usepackage{algpseudocode}%
%\usepackage{listings}%
%\usepackage{listings}%
%\usepackage{newfloat}%
%\usepackage{textgreek}%
%\usepackage[labelfont=bf]{caption}
%\DeclareFloatingEnvironment[name={Fig. S}]{suppfigure}%
%\pagenumbering{gobble}

%\usepackage{dcolumn}% Align table columns on the decimal point
%\usepackage{bm}% bold math
%\usepackage[hidelinks]{hyperref}
%\usepackage[mathlines]{lineno}% Enable numbering of text and display math
% \linenumbers\relax % Commence numbering lines

%\hypersetup{
%    colorlinks,
%    linkcolor = {red!50!black},
%    citecolor = {blue!50!black},
%    urlcolor = {blue!80!black}
%}

%\raggedbottom

\date{}

% No title or authors as per Biophysical reports journal guidelines.

\noindent This Supplementary Information contains additional Figures and Tables as referenced in the main text.

\title{Objective clustering protocol for single-molecule data: A lifetime vs. intensity study}%% For page header:

\author[1]{Michael A.C. Lovemore}
\author[1]{Joshua L. Botha}
\author[1]{Gonfa T. Assefa}
\author[1,*]{Tjaart P.J. Kr\"uger}

\affil[1]{Department of Physics, University of Pretoria, Lynnwood Road, 
Pretoria, 0002, Gauteng, South Africa}

\title{Supplementary Information}

\clearpage
\section*{Supplementary Information}
\addcontentsline{toc}{section}{Supplementary Information}

% Reset counters and redefine figure/table numbering
\setcounter{figure}{0}
\renewcommand{\thefigure}{S\arabic{figure}}

\setcounter{table}{0}
\renewcommand{\thetable}{S\arabic{table}}

%\subsection*{Simulated data additional information}

%\subsection*{Four-state simulation}

\begin{figure}[h!tbp]
    \centering
    
    \includegraphics[width=0.7\linewidth]{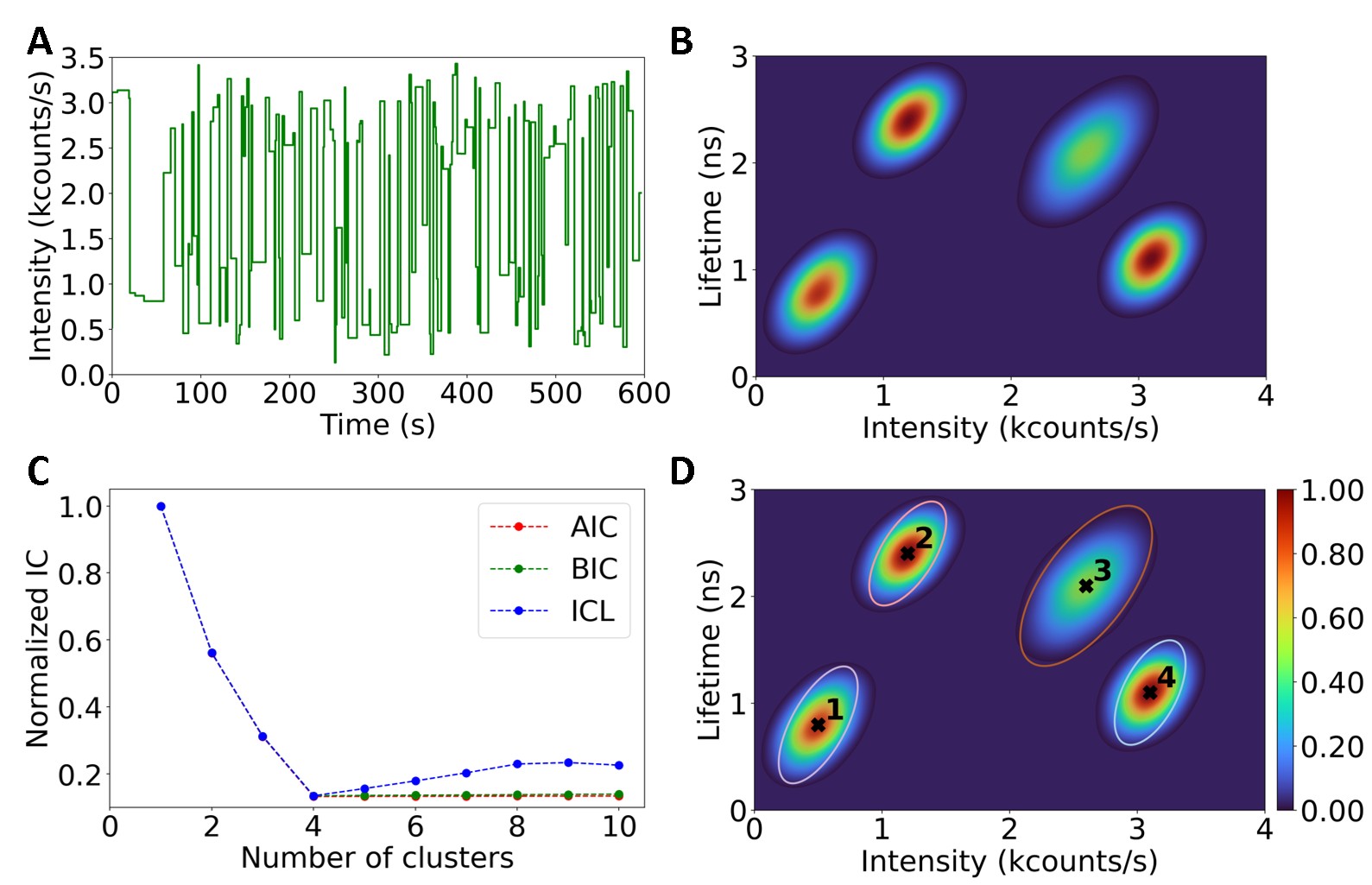}
    \caption{{Clustering results for the four-state simulated lifetime--intensity dataset (300 particles, 10-min traces). A. Representative 40-ms binned intensity trace showing rapid photoblinking. B. Lifetime--intensity distribution of all traces, revealing four populations. C. IC score plots (AIC, BIC, and ICL) identify four clusters as optimal, with mean cluster tightness $\langle D^2\rangle \approx 2$ and FO $\sim 0.05$. D. Clustered distribution with centers at (500 cps, 0.8 ns), (2601 cps, 2.1 ns), (1205 cps, 2.4 ns), and (3100 cps, 1.1 ns) marked by crosses and enclosed by 0.95 confidence ellipses. ARI = 1, confirming accurate recovery of the simulated states.}}

    \label{fourstatesim}
\end{figure}

%\subsection*{Two-state system: Comparing clustering methods}

\begin{figure}[h!tbp]
    \centering
    
    \includegraphics[width=0.7\linewidth]{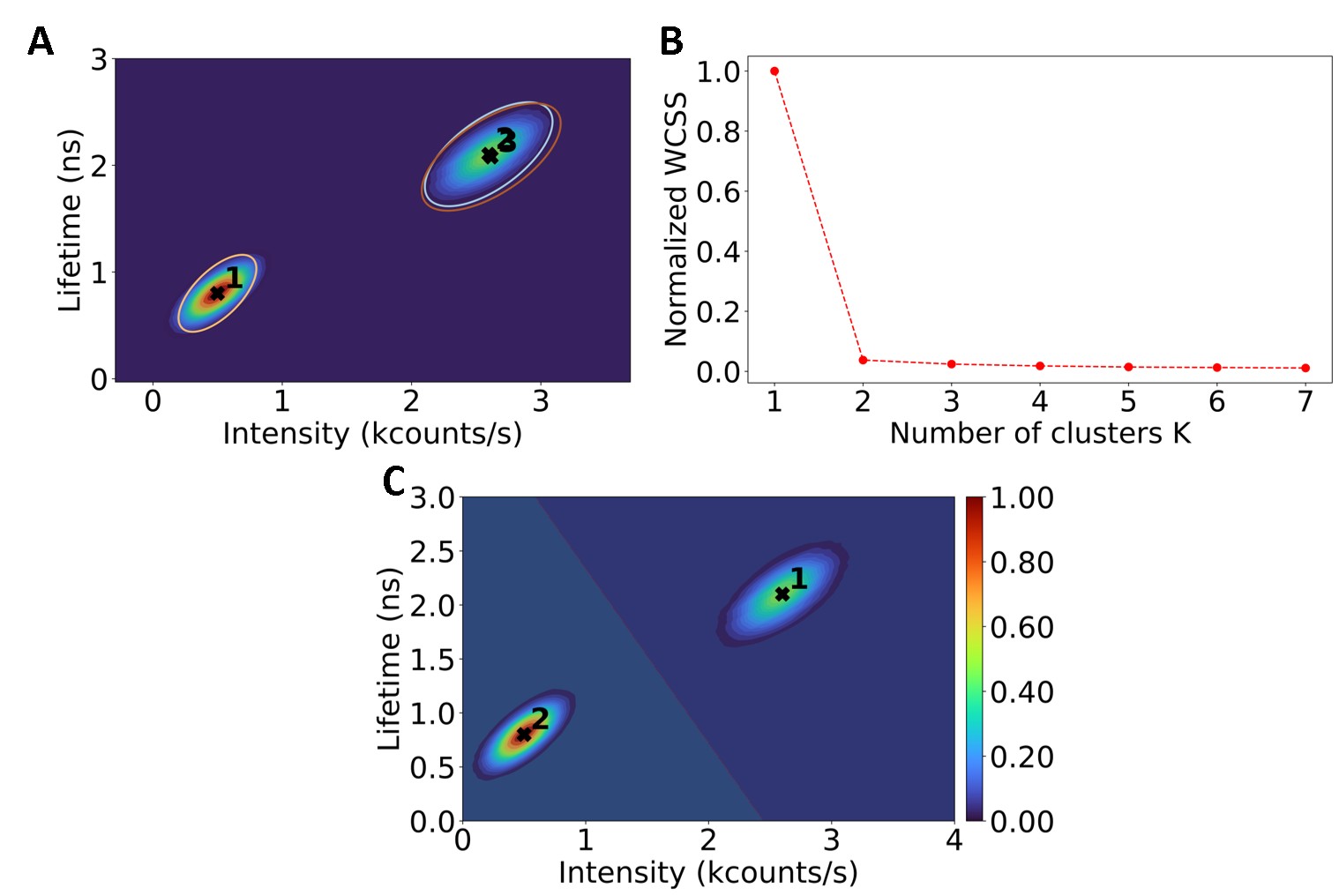}
    \caption{{Comparison of alternative clustering methods on a well-defined two-state simulated fluorescence lifetime--intensity dataset (10-min time, 300 particles). A. DPGMM with three prior components and a Dirichlet concentration of 0.1 assigns two clusters to the same high-intensity population, forming a low-occupancy “ghost” cluster; 0.95 confidence ellipses shown; computation time $\approx 2 \times$ that of the GMM protocol. B. K-means elbow plot identifies two clusters as optimal; computation time $\approx 2 \times$ that of GMM. C. K-means result for two clusters shows a visually reasonable separation of the two populations.}}
    \label{compareclustermethodone}
\end{figure}

%\subsection*{Different two-state blinking models}
\begin{figure}[h!tbp]
    \centering
    \includegraphics[width=0.7\linewidth]{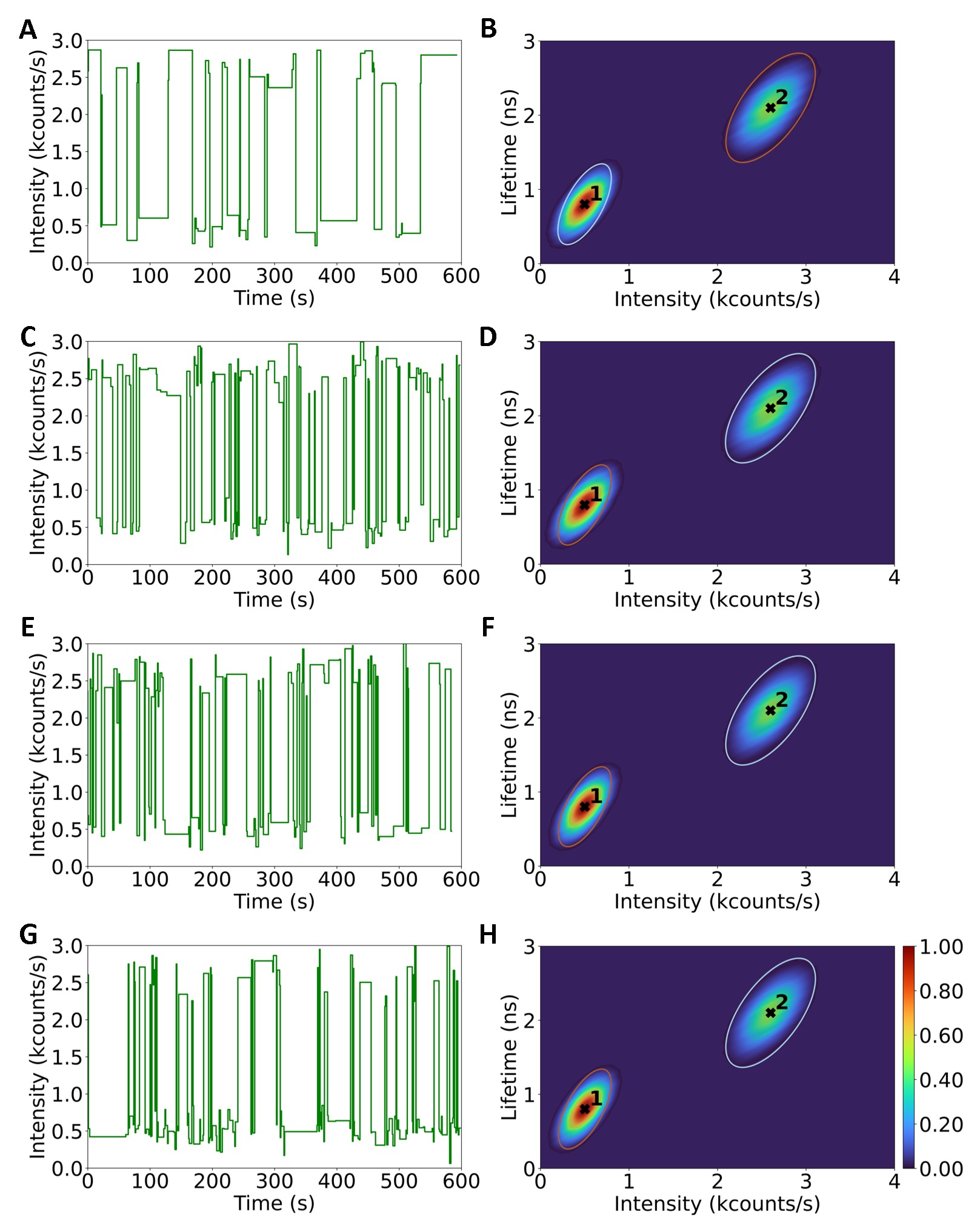}
    \caption{{Clustering outcomes for simulated two-state lifetime--intensity datasets with different blinking models. A, C, E, G: Distributions for simulations with identical mean states but varied on-state dwell times—power-law (A), exponential (C), truncated power-law with $k=1.2$ (E), and truncated power-law with $k=3.5$ (G); off-state dwell times follow a power law in all cases. B, D, F, H: Corresponding clustered solutions with cluster centers (black crosses) and 0.95 confidence ellipses. Mean cluster tightness $\langle D^2\rangle \approx 2$, FO $\sim 0.05$, and ARI = 1, confirming accurate state recovery.}}
    \label{SIotherdwells}
\end{figure}

%\subsection*{Two-state system with additional diffuse states: Comparing clustering methods}

\begin{figure}[h!tbp]
    \centering
    
    \includegraphics[width=0.7\linewidth]{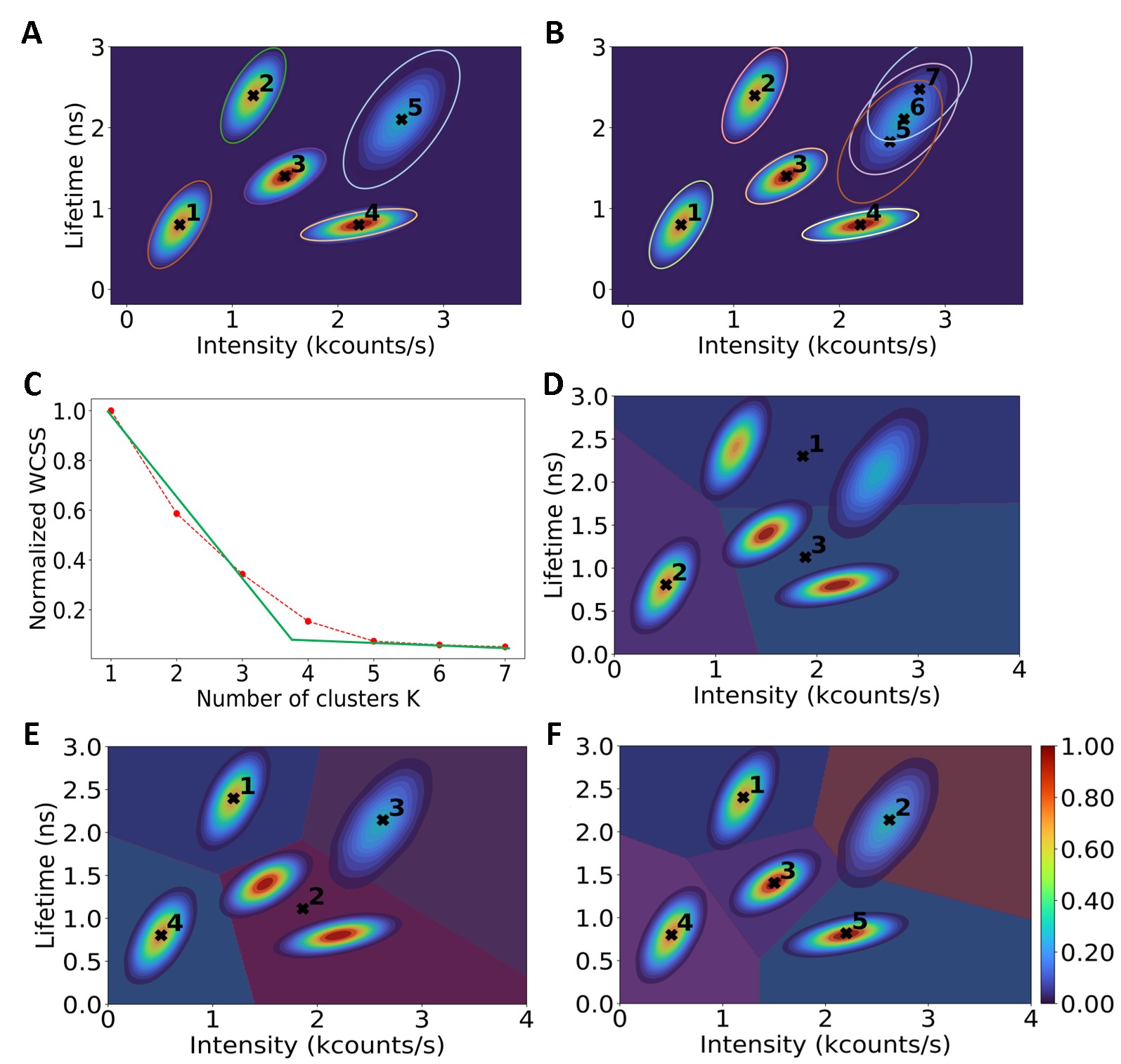}
    \caption{{Comparison of clustering methods on a well-defined five-state simulated dataset (SD = 0.1 $\times$ mean; 10-min time, 300 particles). A. DPGMM with six prior components correctly recovers all five populations. B. DPGMM with eight priors assigns seven clusters, including two low-occupancy “ghost” clusters; 0.95 confidence ellipses shown, and Dirichlet weights set to 0.1. C. K-means elbow plot (red) with green gradient lines fitted to the steep and shallow regimes of the WCSS curve; the elbow is identified at their intersection, where the gradient transitions from rapid to slow decrease, yielding three or four plausible cluster numbers. D, E. K-means results for three (D) and four (E) clusters, showing poor recovery. F. K-means with five clusters explicitly specified, correctly recovering all populations. For both K-means and DPGMM, computation time was $\approx 2 \times$ that of our GMM protocol.}}
    \label{compareclustermethodtwo}
\end{figure}

%\subsection*{Different multi-state blinking models}

\begin{table}[h!tbp]
    \small
    \centering
    \caption{{Means and standard deviations of the simulated intensity and lifetime parameters used to generate the ten-state, multi-state dataset composed of five well-defined states (1--5) and five diffuse, low-probability states (6--10)}. }
    \begin{tabular}{|c|c|c|c|c|}
       \hline
       State & Mean intensity (cps) & Intensity SD (cps) & Mean Lifetime (ns) & Lifetime SD (ns) \\ \hline
       1  & 500  & 120 & 0.8 & 0.22  \\ \hline
       2  & 2600 & 200 & 2.1 & 0.30  \\ \hline
       3  & 1200 & 200 & 2.4 & 0.20  \\ \hline
       4  & 1500 & 200 & 1.4 & 0.20  \\ \hline
       5  & 2200 & 200 & 0.8 & 0.20  \\ \hline
       6  & 1900 & 600 & 1.0 & 0.40  \\ \hline
       7  & 1300 & 600 & 2.0 & 0.40  \\ \hline
       8  & 1000 & 600 & 0.6 & 0.40  \\ \hline
       9  & 3000 & 600 & 1.4 & 0.40  \\ \hline
       10 & 2000 & 600 & 2.2 & 0.40  \\ \hline
    \end{tabular}

    \label{noiseltdsimtable}
\end{table}

\begin{figure}[h!tbp]
    \centering
    \includegraphics[width=1\linewidth]{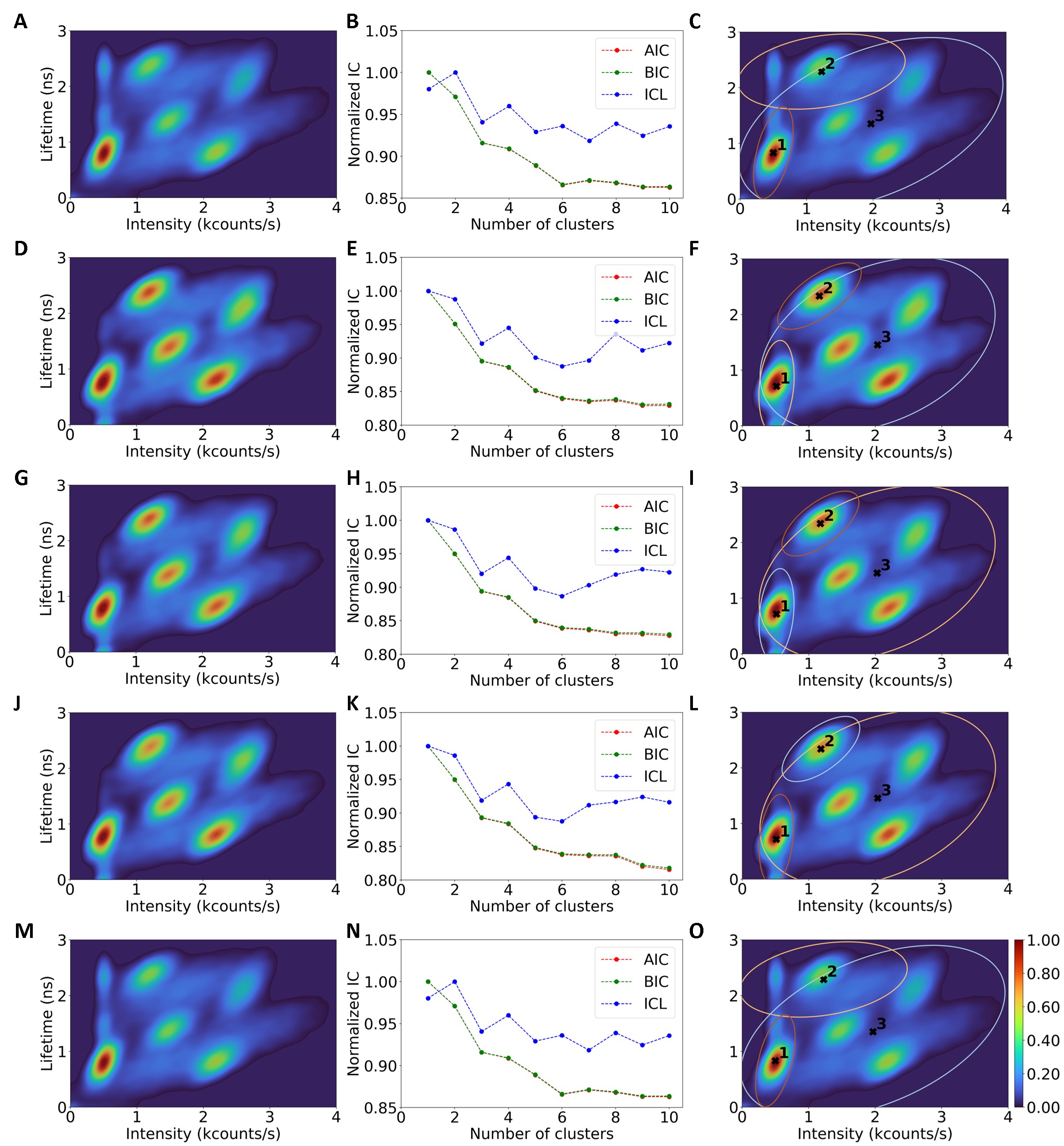}
     \caption{{Clustering results for a ten-state system (five well-defined and five diffuse states) under different blinking models. A--C: power-law on- and off-state dwell times; D--F: exponential on-state dwell times; G--I: truncated power-law on-state dwell times (exponent 1.2); J--L: truncated power-law on-state dwell times (exponent 3.5), with off-state dwell times following a power-law in all cases. Across all blinking models, three clusters are recovered at similar centers (approximately 510, 1170, and 2020 cps with lifetimes of 0.71, 2.3, and 1.45 ns), with mean cluster tightness $\langle D^2\rangle = 1.80$--1.87, FO $\approx 0.03$, and ARI = 0.11--0.12. These results indicate that clustering performance in dense, overlap-dominated datasets is largely insensitive to the choice of blinking model.}}
    \label{blinkingmultistate}
\end{figure}

%\subsection*{Clustering other types of SM data}

\begin{figure}[h!tbp]
    \centering
    \includegraphics[width=0.7\linewidth]{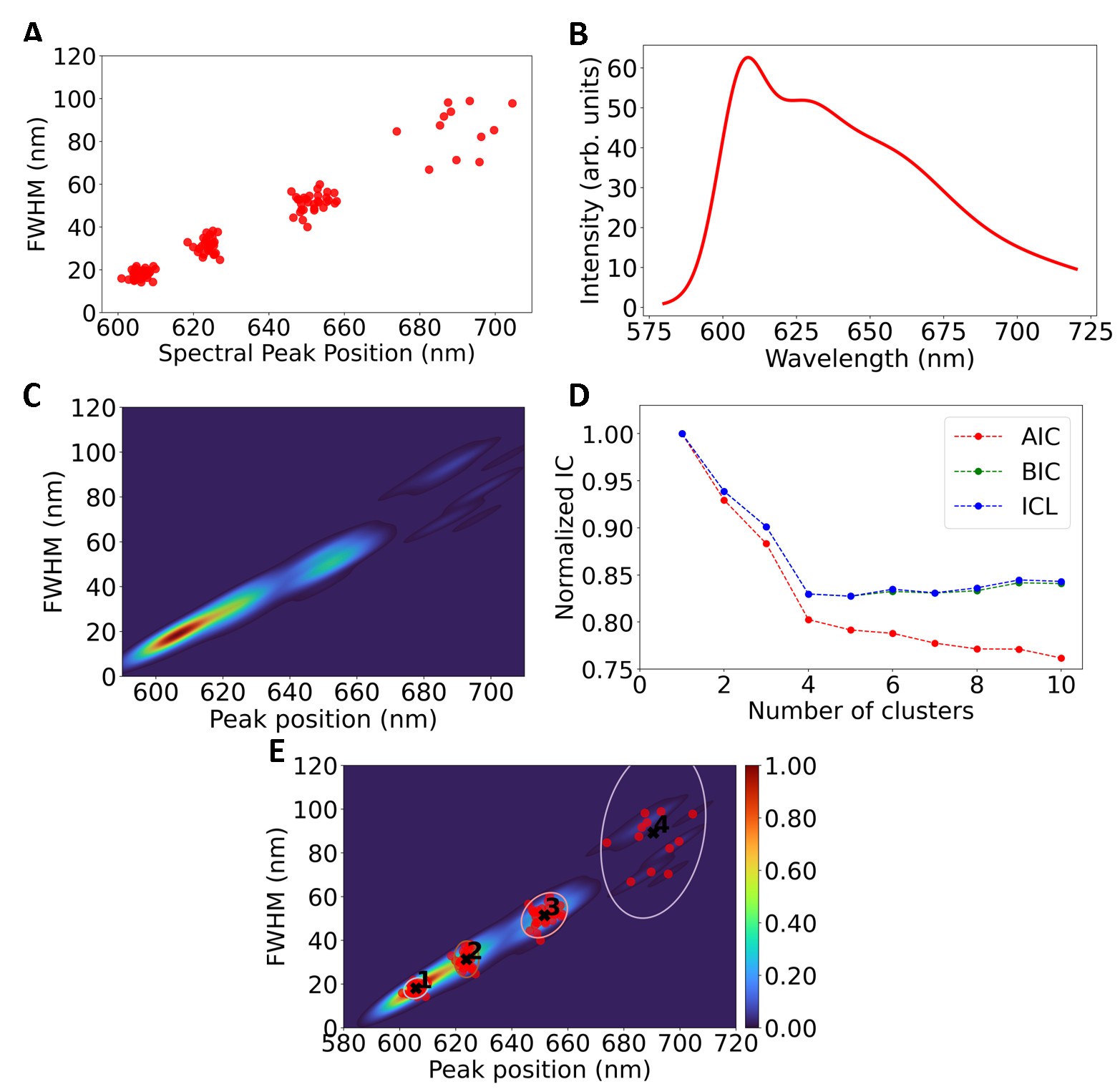}
    \caption{{Clustering of simulated FWHM versus peak-position data for multiple states. A. Scatter plot of the FWHM--peak position distribution. B. Ensemble-averaged intensity versus wavelength; C: Density plot of the FWHM--peak position distribution. D. IC score plots (0.95 confidence) identifying four clusters as optimal. E. Clustered result with centers (black crosses) and 0.95 confidence ellipses. ARI = 1, indicating perfect recovery of the simulated states.}}
    \label{fwhmSI}
\end{figure}

\end{document}